\documentclass[oneside,a4paper,12pt]{article}
\usepackage[utf8]{inputenc}
\usepackage[english]{babel}
\usepackage{amsmath,amssymb,bm}
\usepackage[margin=3cm]{geometry}
\usepackage{mathtools}
\usepackage{graphicx}
\usepackage{float}
\usepackage{wrapfig}
\usepackage{subcaption}
\usepackage{hyperref}

\usepackage{amsthm}

\theoremstyle{definition}

\newtheorem{mydef}{Definition}[section]

\theoremstyle{plain}
\newtheorem{theorem}{Theorem}[section]

\numberwithin{equation}{section}

\begin{document}
%%%%%%%%% TITLEPAGE
\begin{titlepage}

\newcommand{\HRule}{\rule{\linewidth}{0.5mm}} % Defines a new command for the horizontal lines, change thickness here

\center % Center everything on the page
 
%----------------------------------------------------------------------------------------
%	HEADING SECTIONS
%----------------------------------------------------------------------------------------

\begin{tabular}{p{1.5cm} p{12.5cm}}
\vspace{-5pt}
\includegraphics[scale=0.28]{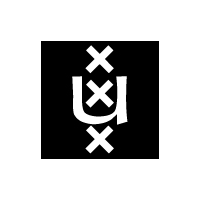}\\[-1.6cm]
&
\vspace{0pt}
\textsc{\Huge University of Amsterdam}\\[1.3cm]

\end{tabular}

\Large MSc Physics\\[0.2cm]
Theoretical Physics Track\\[0.5cm]

\textbf{\Large \\Master Thesis}\\[0.5cm] % Major heading such as course name
%\textsc{\large Minor Heading}\\[0.5cm] % Minor heading such as course title

%----------------------------------------------------------------------------------------
%	TITLE SECTION
%----------------------------------------------------------------------------------------

\HRule \\[0.3cm]
{ \huge \bfseries Violations of the Null Energy Condition in QFT and their Implications}\\[0.4cm]

%{\Large \color{red} \textbf{LHCb SciFi Upgrade}}\\[0.4cm] % Title of your document
\HRule \\[0.3cm]

 \begin{center}
 \textit{{\Large by}}\\[0.4cm]
 \textbf{\Large Dimitrios Krommydas}\\[0.2cm]
 \textbf{\large 10867473}\\

\large \textit{\\ January 2017\\
 60 ECs\\
 September 2015 - January 2017}\\[1.4cm]
 
\vspace{-10mm}
 
 \end{center}
%----------------------------------------------------------------------------------------
%	AUTHOR SECTION
%----------------------------------------------------------------------------------------

\begin{minipage}{0.4\textwidth}
\begin{flushleft} \large
\emph{\\ \Large Supervisor:}\\
\Large Ben Freivogel % Supervisor/Examiner's name
\end{flushleft}
\end{minipage}
~
\begin{minipage}{0.4\textwidth}
\begin{flushright} \large
\emph{\\ \Large Examiner:} \\
\Large  J.P. van der Schaar % Examiner's Name
\end{flushright}
\end{minipage}\\[1.5cm]

% If you don't want a supervisor, uncomment the two lines below and remove the section above
%\Large \emph{Author:}\\
%John \textsc{Smith}\\[3cm] % Your name

%----------------------------------------------------------------------------------------
%	DATE SECTION
%----------------------------------------------------------------------------------------
\Large String Theory Group \\ Cosmology Group\\[0.4cm]
\Large Institute of Physics\\[0.7cm]
%\Large Van der Waals-Zeeman Institute\\[1cm]
%{\large \today}\\[1cm] % Date, change the \today to a set date if you want to be precise

%----------------------------------------------------------------------------------------
%	LOGO SECTION
%----------------------------------------------------------------------------------------
%\includegraphics[scale=1.2]{NIKHEF.png} \hspace{0.5cm} %logos of the institute/company you are working
%\includegraphics[scale=0.3]{lhcb.jpeg} % this will require the graphicx package
 
%----------------------------------------------------------------------------------------

\vfill % Fill the rest of the page with whitespace

\end{titlepage}

\date{}
\let\cleardoublepage
%%%%%%%%%%

\begin{abstract}
	We study violations of the Null Energy Condition (NEC) in Quantum Field Theory (QFT) and their implications. For the first part of the project, we examine these violations for classes of already known and \textit{novel} (first discussed here) QFT states. Next, we discuss the implications of these violations focusing on the example of \textit{Wormhole Traversability}. After reviewing the current literature on the existing restrictions on these violations, we conjecture that NEC violating states are incompatible with the \textit{Semi-Classical Gravity} approximation. We argue that this conjecture provides the only way out of the problems introduced by the violations of NEC in this regime. Building on this, we propose a bound that should hold for \textit{all} QFT states. Finally, we show that both our conjecture and bound hold for some relevant classes of QFT states.

\end{abstract}

\newpage

\tableofcontents
\newpage

\section{Introduction}
\label{intro}
The purpose of this thesis is to explore the violations of the Null Energy Condition (NEC) in Quantum Field Theory (QFT). First, we introduce the Energy Conditions in the framework of General Relativity (GR), their implications and a number of theorems they are
 used to prove (sec. \ref{EC}, \ref{the focusing}, \ref{Penrose Sing}, \ref{GR theorems}). But before even defining the Energy Conditions, it is imperative to understand why there was a need for imposing them. The beauty and the
revolutionary nature of GR can be summarized in one fundamental equation, the Einstein equation\footnote{In truth these are many equations, one for each component of $G_{\mu\nu}$. However, they can be written in this nice compact form.}:

\begin{equation}
\label{Einstein eq}
G_{\mu\nu} \equiv R_{\mu\nu} -\frac{1}{2}R g_{\mu\nu} =8\pi G_N T_{\mu\nu}\;.
\end{equation}
  
Here $G_{\mu\nu}$ is the \textit{Einstein tensor}, a combination of the Ricci tensor $R_{\mu\nu}$ and Ricci scalar $R$ and describes the
 (intrinsic) curvature of spacetime 
 (pseudo-Riemannian manifold). $G_N$ is the Newton's constant and $T_{\mu\nu}$ is 
the \textit{stress energy tensor}, the conserved current of spacetime translations, which describes the matter content of
spacetime. A famous sentence which aspires to explain relation \eqref{Einstein eq} is: \textit{Spacetime tells matter how to move, and matter tells 
spacetime how to curve}. But for all its beauty and intuitive simplicity, the Einstein equation leaves an important question 
unanswered. What properties should matter have, to be considered \textit{physical}? Of course, Einstein already knew 
this when he created his theory. His concerns can be summarized in the following quote \cite{Einstein}: 

 \textit{It [general relativity] is similar to a building, one wing of which is made of fine marble (left part of the equation), 
but the other wing of which is built of low-grade wood (right side of the equation).
The phenomenological representation of matter is, in fact, only a crude substitute
 for a representation which would do justice to all known properties of matter}

An easy way to understand the gravity of the issue is the following:
 without any restrictions on $T_{\mu\nu}$, \textit{any} metric is a valid solution to the Einstein equations. Just pick your favorite metric, 
compute $G_{\mu\nu}$, plug it in the Einstein equations and derive the $T_{\mu\nu}$ for this spacetime. The
 stress energy tensor you calculated will be automatically conserved by the Bianchi identity.
 Thus, to limit the arbitrariness 
of $T_{\mu\nu}$, the \textit{Energy Conditions} are imposed, and it is believed that at least at the classical level, these are conditions 
that the physical matter satisfies. 

And in fact, these restrictions seem to work pretty well. They save us from a lot of awkward situations, a short list of which is
the following: violations of the second law of thermodynamics, violation of ``cosmic censorship'', production of naked singularities,
singularity avoidance inside Black Holes, the breakdown of ``positive mass'' theorems in GR, the breakdown of Penrose Singularity theorem,
 breakdown of Hawking area theorem, the existence of warp drives \& 
 traversable wormholes, which in turn imply faster than
 light travel and closed timelike curves, etc. (sec \ref{GR theorems} ). The traversability of Wormholes (ch. \ref{WH}) is the main implication of the violations of NEC that we will study throughout this thesis.  
 
Since NEC plays such a central role in classical relativistic physics, one might suspect that it is more than a mere assumption. Jan Pieter van der Schaar and Maulik Parikh showed that the NEC can arise in the target space (spacetime) of a bosonic string, from the Virasoro constraint on the world-sheet \cite{Parikh:2014mja}. In later work, M.Parikh et al. showed that NEC can be also derived by assuming the second law of thermodynamics and applying it to the Bekenstein entropy \cite{Parikh:2015wae}. The above findings, suggest that NEC is a fundamental restriction imposed by physical theories and not an ad hoc postulate.

Although all observed forms of classical matter obey at least NEC, this is not the case for all \textit{quantum matter}. There are simple examples of quantum field theory states that violate NEC, even for free field theories in flat spacetime. Some examples are: squeezed vacuum states \cite{Kuo:1993if}, superpositions of the
vacuum and an excited mode (sec. \ref{0+2}), Casimir effect (sec. \ref{Casimir}), moving mirrors \cite{Birrel} and even superpositions
of 1+1 CFT states (sec. \ref{sec:1+1 CFT}). Finally, a notorious example in which NEC violations occur is the Hawking evaporation process \cite{Hawking:1974sw}.

\bigskip

In the first part of this project, we define the energy conditions (sec. \ref{EC}) and we present their form for the popular perfect fluid example. We then derive the \textit{Raychaudhuri equations} (sec. \ref{Raychaudhuri equations}), which combined with the Einstein equations and NEC, lead to a formulation of the \textit{Null Focusing Theorem} (sec. \ref{the focusing}). Next, we use the null focusing theorem to interpret the \textit{Penrose Singularity Theorem} (sec. \ref{Penrose Sing}), which is presented along with a long list of GR results (sec. \ref{GR theorems}) in order to highlight the importance of NEC.

Another important aspect of NEC in GR is that it protects us from exotic spacetimes which allow for causality violations, time-travel, and other troublesome effects. We dedicate ch. \ref{WH} to the detailed analysis of one of these spacetimes, the \textit{Traversable Wormhole}. We first present the well-studied case of the non-traversable wormhole which resides in the maximally extended Schwarzschild spacetime (sec. \ref{Schwarzschild Non-Traversable Wormhole}). Then, we analyze the features of the famous, traversable \textit{Ellis Wormhole} (sec.\ref{Ellis Traversable Wormhole}) and finally use NEC to show that this solution is not allowed in GR.

 As we mentioned, QFT does not always respect NEC. In ch. \ref{Viol}, we provide evidence for this lack of respect. We first present one of our results; a general class of 1+1 CFT states which, as we explicitly show, leads to violations of NEC (sec. \ref{sec:1+1 CFT}). We then reproduce a computation of \cite{Kuo:1993if}; namely, we calculate the expectation value of the relevant components of $\hat{T}_{\mu\nu}$ in the vacuum plus two modes state, and finally show that it violates a quantum version of NEC. After citing the results for the expectation value of $\hat{T}_{\mu\nu}$ for a free scalar in the Casimir state, we show how it violates NEC and explain the significance of the directionality of the Casimir plates. In the end of ch. \ref{Viol}, we display our attempt to violate NEC for a general class of states which are close to the vacuum (sec. \ref{perturbed}); i.e. $e^{-i\int J(x)d^{d}x\Phi(x)}\lvert 0 \rangle$ where $J$ is an arbitrarily small, real source and $\Phi$ a massless free scalar field. Surprisingly, we see that no such violation occurs.

In ch. \ref{Constraints}, we discuss a collection of the most relevant local and non-local constraints on the violations of NEC. First, we review an averaged version of a quantum null energy condition, called the \textit{Averaged Null Energy Condition} (ANEC) and mention which GR results can be salvage by imposing it in the quantum regime (sec. \ref{ANEC}). Next, we discuss the \textit{Quantum Inequalities} (QI's), a set of bounds on the magnitude of the NEC violating flux which are known to hold for \textit{arbitrary} quantum states and \textit{all} inertial observers in Minkowski spacetime. More interestingly, QI's have been shown to hold also for a number of curved spacetimes. We close ch.\ref{Constraints}, with a review of some of the most recent developments in the area of quantum energy conditions, namely the \textit{Quantum Null Energy Condition}  (sec. \ref{QNEC}) and the \textit{Quantum Focusing Conjecture} (sec. \ref{QFC}), introduced by R. Bousso, A. Wall et al. The Quantum Focusing Conjecture is a proposal for a quantum corrected version of the \textit{Classical Focusing Theorem} (sec. \ref{the focusing}). The quantum null energy condition, which is a pure QFT statement, arises naturally by imposing this conjecture on a specific setup; moreover, it was proven in separate work \cite{Bousso:2015wca}, the same year when the quantum focusing conjecture paper \cite{Bousso:2015mna} appeared. In a nutshell, the quantum null energy condition says that the quantum stress energy tensor is bounded from below by the second derivative of some entanglement entropy.

In ch. \ref{Semiclassical}, we study the violations of NEC in semi-classical gravity. Since we do not  have a complete theory of Quantum Gravity yet, it is of utmost importance to understand these violations in a regime of physics over which we have more control. After a brief introduction to semi-classical gravity approximation (ch. \ref{Semiclassical}), we discuss its regime of validity (sec. \ref{VSCG}). We explain that there are cases, even in small energy scales compared to the Planck scale, where the semi-classical approximation cannot be trusted. For instance, large fluctuations of the metric, induced by large $\hat{T}_{\mu\nu}$ fluctuations, break the fixed background assumption. In the next section (sec. \ref{conjecture}), we conjecture that \textit{states for which a quantum version of NEC is violated, lead to such a breakdown}. Hence, at the semi-classical level NEC still holds, allowing us to generalize our GR theorems and exclude ``exotic'' solutions. In sec. \ref{Very QM}, we give a simplified version of these very quantum states and present some intuition as to why these states are incompatible with the semi-classical regime. We then suggest, that if one wishes to make precise statements about the fluctuations of $\hat{T}_{\mu\nu}$, he/she should study the probability distribution function of the operator in the state of interest. Since finding all eigenvalues and their weights is usually an impossible task, one should find and study the relevant features of the distribution (sec. \ref{prob}). As we find out, a remarkably suitable quantity for this assignment is the \textit{variance} of the distribution. Hence, we introduce a dimensionless quantity which contains the variance and is suitable for quantifying the ``quantumness" of a state. We then use this quantity to test the validity of our conjecture, for some relevant classes of states (sec. \ref{test conj}). In the end of this chapter (ch. \ref{Semiclassical}), inspired by our previous analysis, we propose a QFT bound (sec. \ref{proposal}) which we believe should hold for all QFT states. Our bound, namely:

\begin{equation*}
\langle T^{s}_{\mu\nu} K^\mu K^\nu \rangle + \sigma \geq 0,
\end{equation*}
involves the sum of the expectation of the null projected component of a smeared stress energy tensor and the standard deviation. The smearing is important, since it introduces a natural cut-off to the UV (short distance) divergences present in $\langle \hat{T}_{\mu\nu}(x) \hat{T}_{\rho\sigma}(x) \rangle$ and transforms the point-wise defined stress tensor into a quantity more appropriate for describing physical measurements (sec. \ref{Smear}). In the end of sec. \ref{proposal}, we explicitly show that our bound holds for some interesting classes of states.

% from hereon we put $\hbar=c=1$

\newpage

\section{Null Energy Condition}
\label{nec}

In this chapter, we study the \textit{Null Energy Condition (NEC)} in GR. We start by defining it, along with other relevant \textit{Energy Conditions} in sec. \ref{EC}, and then discuss its role in the formulation and proof of many important GR theorems (sec. \ref{the focusing}, \ref{Penrose Sing}, \ref{GR theorems}). Since the first appearance of the energy conditions, the set of matter distributions/spacetimes that are considered physical has been greatly enlarged. As a result, the only remaining local condition that is believed to hold today is NEC. Fortunately, this last stronghold is known to hold for all observed classical matter.

\subsection{Energy Conditions}
\label{EC}

The energy conditions are coordinate-invariant restrictions on the \textit{Stress-Energy Tensor}. This invariance is enforced 
 by requiring that the quantities involved are scalar. The scalar quantities which are usually considered, are contractions of $T_{\mu\nu}$ 
 with timelike $t^\mu$ or null vectors $k^\mu$. Typically, these conditions are discussed for the case of the perfect fluid. For the purposes of this introductory chapter, we continue this tradition. For a perfect fluid source, $T_{\mu\nu}$ is:

\begin{equation}
T_{\mu\nu}=(\rho+p)U_{\nu}U_{\mu}+pg_{\mu\nu}\;.
\end{equation}

Where $U_\mu$ is the fluid four-velocity, $g_{\mu\nu}$ is the spacetime metric, $\rho$ is the energy density and $p$ is the pressure 
 in the $x^i$ direction ($i=1,3,4$). Without further ado, we present a list of classical, local energy condition following Carroll \cite{Carroll:2004st}

\begin{mydef} 
Null Energy Condition: $T_{\mu\nu} k^\mu k^\nu \geq 0$ for all null vectors $k^\mu$. 
\end{mydef}
 
 As we shall see shortly, we started with the weakest (in the sense that it is implied by all the others) among these local energy conditions. However, our intention is to discuss violations of these conditions. So the less restrictive the condition is, the harder it is to violate. Hence, NEC is the hardest to violate, which justifies this seemingly backward order.  
 
For a perfect fluid, NEC translates to $\rho+ p \geq 0$, and is implied
by the stronger Weak Energy Condition.

\begin{mydef} 
Weak Energy Condition (WEC): $T_{\mu\nu} t^\mu t^\nu \geq 0$ for all timelike vectors $t^\mu$. 
\end{mydef}

When $p$ is isotropic, WEC holds for all timelike vectors if $T_{\mu\nu} U^\mu U^\nu \geq 0$ \textit{and} 
$T_{\mu\nu} k^\mu k^\nu \geq 0$ for some null vector. For a perfect fluid:

\begin{equation}
T_{\mu\nu} U^\mu U^\nu = \rho,  
\end{equation}

\begin{equation}
T_{\mu\nu} U^\mu U^\nu = (\rho + p)(U_\mu k^\mu)^2.  
\end{equation}

 WEC implies $ \rho \geq 0$ and $ \rho + p \geq 0$. When is worried about non-physical matter, WEC looks like the first thing he/she should consider since it straightforwardly excludes negative energy from all points in spacetime. However, as we mentioned, there are matter distributions of spacetimes we consider physical that \textit{do} violate it. A renown example is AdS spacetime, which is filled with negative energy\footnote{One need not worry about AdS violating NEC since the negative energy density is compensated by positive pressure.}.

\begin{mydef} 
Dominant Energy Condition (DEC) : $T_{\mu\nu} t^\mu t^\nu \geq 0$ (WEC) and  $T_{\mu\nu} T^\nu_{\phantom{\lambda}\lambda}  t^\mu t^\lambda \leq 0$, 
for all timelike vectors $t^\mu$.
\end{mydef}

The second part of the definition is the requirement that $T_{\mu\nu} t^\mu$ is a non-spacelike vector. For a prefect fluid, this
translates to $\rho \geq \vert p \rvert$, i.e. the energy density must be non-negative and greater or equal than the magnitude of the
pressure.

\begin{mydef} 
Null Dominant Energy Condition (NDEC) : $T_{\mu\nu} k^\mu k^\nu \geq 0$ (NEC) and  $T_{\mu\nu} T^\nu_ {\phantom{\lambda}\lambda} 
 k^\mu k^\lambda \leq 0$, for all null vectors $k^\mu$.
\end{mydef}

The second part of the definition is the requirement that $T_{\mu\nu} k^\mu$ is a non-spacelike vector. NDEC excludes all sources excluded by DEC, apart from those with negative vacuum energy, i.e. $\rho=-p$. 

\begin{mydef} 
Strong Dominant Energy Condition (SEC) : $T_{\mu\nu} t^\mu t^\nu \geq \frac{1}{2} T^\lambda_ {\phantom{\lambda}\lambda} t^\sigma
 t_\sigma $, for all timelike vectors $t^\mu$.
\end{mydef}

This implies that $ \rho + p \geq 0$ and $ \rho + 3p \geq 0$. In this case SEC implies NEC and restricts 
negative pressure from being too large.
We should point out that these conditions are very strong, in the sense that they are local. This means that they should hold
for \textit{every point} in spacetime and are, thus, far more restricting than their global counterparts, the \textit{Averaged} energy
conditions, which we shall discuss in a later chapter (ch. \ref{ANEC}).

%++ Pfenning stuff

%++ maybe diagrams?

\bigskip

\bigskip

\subsection{Raychaudhuri Equations}
\label{Raychaudhuri equations}

Suppose that one wants to follow the evolution of a collection of bodies 
(particles) falling under their own gravity. If no 
interaction is assumed, he/she would expect them to attract and tend to converge. A way to formulate this evolution in the context of
general relativity is by the use of the \textit{Raychaudhuri Equation} \cite{Raychaudhuri}. For the following discussion one needs to define the notion of a
\textit{congruence}:

\begin{mydef}
 
Congruence: a set of curves in an open region of spacetime, such that every point in the region lies on precisely one curve.

\end{mydef}

%++ picture? 

If the set of curves is restricted to that of geodesics, then one obtains a \textit{geodesic congruence}. It is convenient to think of a geodesic congruence as 
a set of paths followed by non-interacting particles moving through spacetime. If their geodesics cross, a
\textit{caustic} is formed, i.e. the congruence comes to an end.

The starting point of the derivation is defining a vector field $U^\mu=dx^\mu/d\tau$, tangent to a timelike geodesic congruence. Then, one has to define a separation vector 
$V^\mu$, which points from one geodesic of the congruence to another in its vicinity. For a general congruence this vector obeys:

\[\frac{DV^\mu}{d\tau} \equiv U^\nu \nabla_\nu V^\mu = B^\mu_{\phantom{\nu}\nu} V^\mu, \quad B^\mu_{\phantom{\nu}\nu} = \nabla_\nu U^\mu. \]

\begin{figure}[H]
\centering
\includegraphics[width=0.3\textwidth]{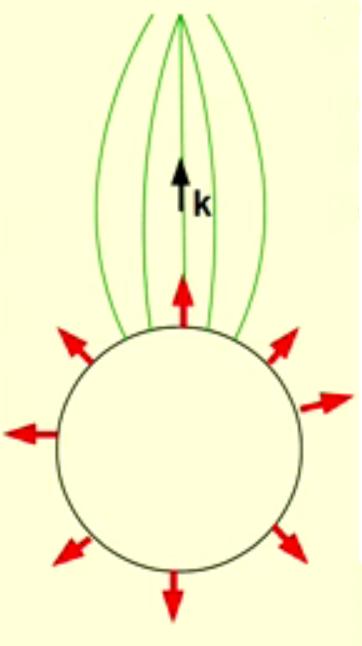}
\caption{ (Taken from \cite{Wall talk}) A 2-D picture of the geodesic sphere. k is a null vector along a null geodesic, and the green curves portray both the focusing and shearing of the null congruence. }
\label{congruence}
\end{figure}

Hence, $B^\mu_{\phantom{\nu}\nu} $ tells as how much $V^\mu$ fails to be parallelly transported along the congruence. In other words, it describes how far 
the geodesics are from being parallel. The picture (fig. \ref{congruence}) one should have in mind, is a sphere of test particles centered at 
a point. At the end of the day, what one wants to describe is the evolution of the particles with respect to their central geodesic. %Thus we will take three
%normal vectors, defined as $V^\mu$  orthogonal to the congruence and follow their evolution.
A quantitative description
 of this evolution can be obtained by following solely the behavior of $B^\mu_{\phantom{\nu}\nu}$.

At each point $p$ a subspace of the tangent space $T_p M$ is considered, which corresponds to vectors normal to the timelike vector field
$U^\mu$. Any vector can be projected to this subspace with the help of a projection tensor 
$P^\mu_{\phantom{\nu}\nu} = \delta^\mu_{\phantom{\nu}\nu} + U^\mu U_\nu$. It is easy to see, using $U^\mu U_\mu = -1 $ (timelike norm) and $U^\mu \nabla_\mu U^\nu = 0 $ (geodesic), that
$B^\mu_{\phantom{\nu}\nu}$ belongs to  $T_p M$, i.e. $U^\mu B_{\mu\nu} = U^\mu \nabla_\nu U^\nu = 0 $ and 
$U^\nu B_{\mu\nu} = U^\nu \nabla_\nu U^\nu = 0 $. At this point, it is useful to decompose $B_{\mu\nu}$  into a symmetric and an anti-symmetric part
 $B_{\mu\nu}= B_{(\mu\nu)} + B_{[\mu\nu]}$. $B_{(\mu\nu)}$ can be further decomposed into a trace and a trace-free part. Finally one should notice that
 since $B_{\mu\nu}$ belongs to the subspace of vectors normal to $U^\mu$, the projection tensor operator can be used to take traces. When the dust settles, one has:

\[B_{\mu\nu}= \frac{1}{3}\theta P_{\mu\nu} + \sigma_{\mu\nu} + \omega_{\mu\nu}\;.\]

The quantity of interest here is the scalar $\theta$ (trace of $B_{\mu\nu}$) which is the \textit{expansion} of the congruence, or 
in our sphere picture, the expansion of the sphere with center the geodesic we considered the deviation from. $\sigma_{\mu\nu}$ is the 
\textit{shear} tensor. It describes the rate by which the geodesic ball shears into an ellipsoid without any change in volume, and it is a symmetric (0,2) tensor. This symmetry represents the fact that our sphere gets
 distorted in a symmetric way. The last constituent of the equation is the \textit{rotation/vorticity}, again without any change in the 
volume of the sphere, and it is the anti-symmetric part of $B$. The anti-symmetry accounts for the usual clockwise and anti-clockwise 
options. 

Finally, we can write down an equation which describes the evolution of our congruence. To obtain it, we take the covariant 
derivative of $B_{\mu\nu}$ along $U^\mu$:

\begin{equation}
\label{ParB}
\frac{DB_{\mu\nu}}{d\tau}\equiv U^\lambda \nabla_\lambda = -B^\lambda_{\phantom{\nu}\nu}B_{\mu\lambda} -R_{\sigma\mu\nu\lambda} U^\sigma U^\lambda.   
\end{equation}

Taking the trace of \eqref{ParB} we obtain the \textit{Raychaudhuri equation} for the expansion:
 
\begin{equation}
 \frac{d\theta}{d\tau} = -\frac{1}{3}\theta^2 - \sigma_{\mu\nu} \sigma^{\mu\nu} + \omega_{\mu\nu} \omega^{\mu\nu} - R_{\mu\nu} U^\mu U^\nu.
\end{equation}

\subsubsection{Raychaudhuri Equation for Null Geodesics}

Since we are interested in studying NEC, the relevant process for us is the evolution of \textit{null} congruences. % for reasons that will become clearer in \ref{Flaring} 
 Hence here, we derive the Raychaudhuri 
equation for such congruences, which at first sight seems incompatible with the above derivation. The problem one faces is that since now the tangent vector 
$K^\mu= \frac{dx^\mu}{d\lambda}$ is a null vector, the subspace of vectors normal to it will contain itself. What we should study in this case is the evolution of 
spatial vectors in a two-dimensional space normal to $K^\mu$; since there is no unique way to define this subspace, we will need to define an auxiliary null 
vector which points in the opposite direction and hope that it will not show up in our final equation. The vector is chosen so that it has the following 
normalization:

\begin{equation}
l^{\mu} l_{\mu} =0, ~ l^{\mu} K_{\mu} =-1.
\end{equation}

Next, we write the projection operator\footnote{ metric on the tangent space to our vectors} such that it kills $K^\mu$ and $l^\mu$, and projects everything else onto 
the tangent subspace to these vectors:

\begin{equation}
 h_{\mu\nu}= g_{\mu\nu} + K_{\mu}l_{\nu} + K_{\nu}l_{\mu}\;.
\end{equation}

Again the failure of the deviation vector $V_{\mu}$ to be parallelly transported, is described by a tensor $B^\mu_{\phantom{\nu}\nu} = \nabla_\nu K^\mu$. However, the said tensor contains too much information for our purposes, so the quantity we consider is:

\begin{equation}
 B'^\mu_{\phantom{\nu}\nu} = h^\mu_{\phantom{\alpha}\alpha}h^\beta_{\phantom{\nu}\nu}B^\alpha_{\phantom{\beta}\beta}\;.
\end{equation}

Finally, to obtain the Raychaudhuri equation, we proceed as in the timelike case:

\begin{equation}
\label{Null Raych}
 \frac{DB'_{\mu\nu}}{d\tau}\equiv K^\lambda \nabla_\lambda = -B^\lambda_{\phantom{\nu}\nu}B_{\mu\lambda} 
 -h^\alpha_{\phantom{\mu}\mu}h^\beta_{\phantom{\nu}\nu}R_{\sigma\alpha\beta\lambda} U^\sigma U^\lambda.
\end{equation}

After taking the trace of \eqref{Null Raych}, we obtains:

\begin{equation}
\label{RaychL}
\frac{d\theta'}{d\lambda}= -\frac{1}{2}\theta'^2 - \sigma'_{\mu\nu} \sigma'^{\mu\nu} + \omega'_{\mu\nu} \omega'^{\mu\nu} - R_{\mu\nu} K^\mu K^\nu,
\end{equation}

 % ++ (5) Tells us how the rate of the expansion of the area of a null congruence evolves per unit area, along a curve parametrized by an affine parameter $\lambda$.

%\begin{equation} \theta=lim..  \frac{1}{\mathcal{A}} \frac{dA}{d\lambda} \end{equation}

% where $\mathcal{A}$ is an infinitesimal area

which is the desired equation that describes the evolution of the null expansion. Finally we note that our expression is indeed independent of $l_{\mu}$, as required.
 
\begin{figure}[H]
\centering
\includegraphics[width=0.8\textwidth]{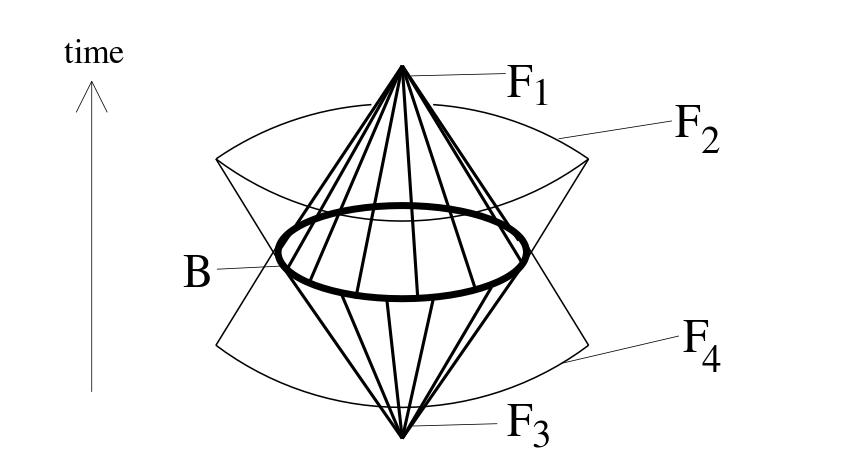}
\caption{(Taken from \cite{Bousso:2002ju})\textit{Causal diamond on a 4-D Minkowski spacetime, where one spatial dimension has been omitted}: starting with a null congruence B on a fixed time slice, we draw the light-cones at each point. Four null hypersurfaces arise, from which $F_1$ and $F_3$ form the causal diamond. The expansion of B along hypersurface $F_1$ is negative (ingoing rays) and along $F_2$ positive (outgoing rays). } 
\label{Null congruence}
\end{figure}

\bigskip

\bigskip

\subsubsection{The focusing Theorem}
\label{the focusing}
%++
We are free to choose the vector $U^\mu$ to be hypersurface orthogonal, such that $\omega_{\mu\nu}=0$. $\sigma'_{\mu\nu} \sigma'^{\mu\nu} \geq 0$ since it is a 
spatial tensor. Notice that we have gotten this far without the using the Einstein equation, so up to this point, the
derivation was
very general.
To extract results on how such a congruence \textit{should} behave in the presence of gravity, we now make use of the Einstein equations:

\begin{equation}
\label{Einstein equations}
G_{\mu\nu} = 8\pi G_N T_{\mu\nu}\;.
\end{equation}

And after rewriting them in a form that suits our purposes, we contract with the null vector:

\begin{equation}
\begin{split}
R_{\mu\nu} &= 8\pi G (T_{\mu\nu}- \frac{1}{2} T g_{\mu\nu}) K^\mu K^\nu \\
&= 8\pi G_N T_{\mu\nu}\;.
\end{split}
\end{equation}

By imposing NEC $T_{\mu\nu} k^\mu k^\nu \geq 0$, and plugging everything back to \eqref{Einstein equations}, we obtain:

\begin{equation}
\label{focusing equation}
 \frac{d\theta}{d\lambda} = -\frac{1}{2}\theta^2 - \sigma_{\mu\nu} \sigma^{\mu\nu} - T_{\mu\nu} K^\mu K^\nu \leq 0.
\end{equation}

What above result tells us, is that \textit{the rate by which the expansion of our congruence changes is always negative}. If we go back to 
our sphere picture and think about what happens to the null rays emanating from it, we will see that they always try to focus. If we start with $\theta \leq 0$, then the light rays will keep focusing. If we start with $\theta > 0$, then the congruence will either become stationary ($\theta = 0$) and then start focusing
or will reach infinity before that happens (depending on our spacetime). This is what we would normally expect from gravity to do. Thus here,
the NEC translates into the statement that gravity is \textit{attractive}. It is useful to formulate the above observations into a theorem:

\begin{theorem}
\label{FocThe}
 The classical (Null) focusing theorem: combining the Raychaudhuri equation, the Einstein equation and NEC we obtain that the evolution of 
the expansion for a null congruence is always $\frac{d\theta}{d\lambda}\equiv\theta' \leq 0$, thus null rays always tend to focus.
\end{theorem}

\bigskip

\bigskip

\subsection{Penrose Singularity Theorem}
\label{Penrose Sing}

\bigskip

The Penrose Singularity Theorem is regarded by many as one of the first genuine post-Einsteinian results of GR. It was derived by R. Penrose in 1965 \cite{Penrose Th}, the same year the CMB was measured, miraculously 50 years after the derivation of the Einstein equations. The original purpose of Penrose was to prove that deviations from spherical symmetry are not able to prevent the formation of singularities, as previously suggested by Oppenheimer \cite{Oppenheimer:1939ue}. 
 
To achieve this goal, Penrose introduced the concepts of \textit{incompleteness} and closed \textit{trapped surface}, which now play a central role in the relativistic world. Without further ado, let us define these quantities, and introduce the theorem. In the end of the section, we provide some physical intuition as to what it means.

The first thing needed for the proof is a sharp definition of a \textit{singularity}. There are well-known examples in which the most obvious definition as ``the point where some curvature scalar diverges'', fails\footnote{Examples are the Taub-NUT singularities, which are formed by incomplete geodesics spiraling infinitely around a topologically
closed spatial dimension, while all curvature scalars are finite \cite{Misner:1965zz}.}. A universal definition is the following:  

\begin{mydef}
\label{sing def}
A singularity in a Lorentzian manifold is an incomplete, endless curve \cite{Senovilla:2014gza}.
\end{mydef}

Nearly all singularity theorems, after the Penrose one, prove the existence of ``geodesic incompleteness''. Another crucial notion is that of a closed trapped surface. Imagine that you have a 2+1 dimensional spacetime. Pick a spacelike surface (Cauchy slice), draw a circle (congruence) and the light-cones at each point of the circle. In your drawing, there are four null hyper-surfaces emanating from the circle, two towards the future and two towards the past. Drawing another spacelike surface, you can follow the evolution of the circle in time. On one of each of these future/past null surfaces the circle gets smaller and on the other is getting bigger (assuming that spacetime is stationary). A null future/past-trapped surface is a circle, which contracts on both future/past hyper-surfaces. This is exactly what happens on every null congruence inside a Schwarzschild horizon. An intuitive example of this behavior is described in the following fig. \ref{penrose surface} .

\bigskip

\begin{figure}[H]
\centering
\includegraphics[width=\textwidth]{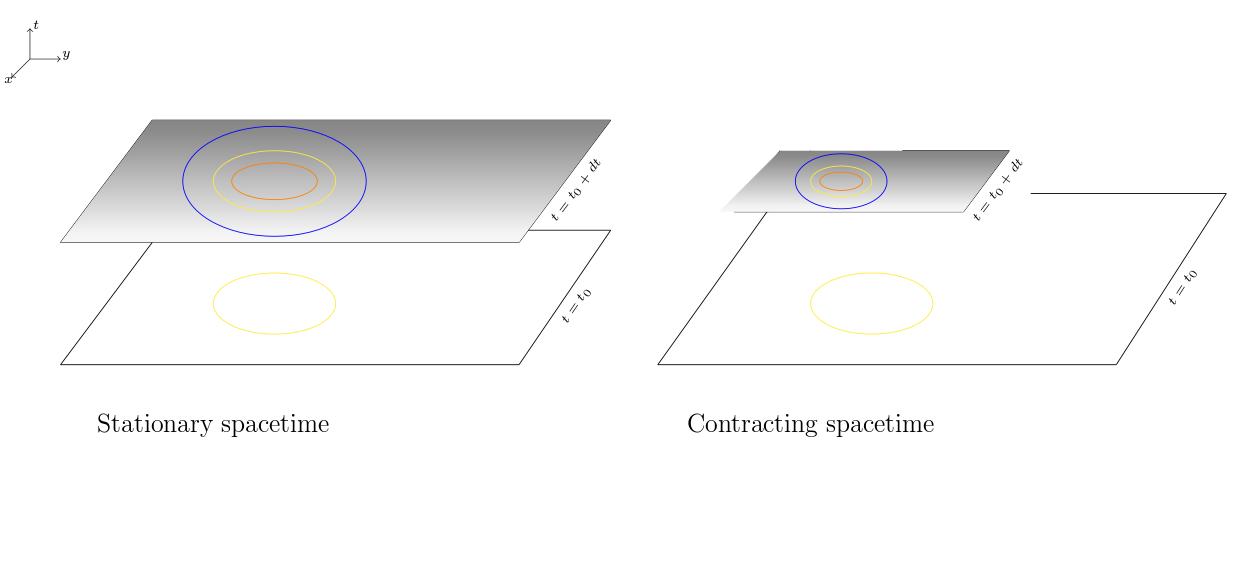}
\caption{ (Taken from \cite{Senovilla:2014gza}) \textit{The evolution of a Null congruence in 4-D spacetimes, where one spatial dimension is omitted}: the first diagram depicts this evolution for a stationary spacetime. An initial null congruence (yellow circle) will either expand (blue circle) or contract (orange circle), depending on whether light is ingoing or outgoing (fig. \ref{Null congruence}). However, if spacetime is contracting fast enough, the congruence on a later slice will contract in both the ingoing and outgoing case.}
\label{penrose surface}
\end{figure}

\bigskip

A \textit{closed trapped surface} is in addition compact, without boundary. Having acquired the required knowledge, we present the Penrose Singularity Theorem: 

\begin{theorem}
\label{PST}
If the space-time contains a non-compact Cauchy hyper-surface $\Sigma$ and a closed future-trapped surface, and if the Null Energy 
Condition holds for null $K^\mu$, then there are future incomplete null geodesics.
\end{theorem}

The assumption of having a well-defined Cauchy evolution problem, goes also by the name \textit{global hyperbolicity}. It is the requirement that there are no timelike nor null closed curves and that spacetime does not have any holes through which information can go in and out off. Furthermore, one can interpret the non-compactness as the requirement that space is infinite.

Using this new intuition and the focusing theorem (th. \ref{FocThe}) implied by NEC, it is almost already clear that a singularity, in the Penrose sense, will form. We know that a closed future-trapped surface will form eventually, and the only problem we need to solve is that our geodesics may cross (form a caustic) instead of terminating on a singularity. This problem is solved by assuming \textit{global hyperbolicity}, which implies that the topology of spacetime cannot change. Thus, at least one geodesic terminates and cannot be extended any further.

\subsection{Other GR results}
\label{GR theorems}
  In previous section (sec. \ref{Penrose Sing}), we focused on explaining the Penrose Singularity Theorem and described the role of NEC in its formulation. To make the importance of NEC even more pronounced, we cite a detailed but still not exhaustive list of results in GR, for which NEC is a central assumption \cite{Curiel:2014zba}.

\begin{enumerate}

\item Singularity formation, after a gravitational collapse, in spatially open spacetimes \cite{Penrose Th}. 

\item Singularity formation, in  asymptotically flat  spacetimes  with a non-simply  connected Cauchy surface \cite{Gannon} \cite{Lee}.

\item Event horizon formation, after gravitational collapse \cite{Penrose Th} \cite{Penrose 1969}.

\item Apparent horizons, trapped and marginally trapped surfaces must reside in asymptotically flat black holes \cite{Wald 1984}.

\item Second Law of black-hole mechanics \cite{Hawking 1971}. 

\item Second Law of generalized black-hole mechanics %\footnote{The area of a generalized black hole always increases.} \cite{Hayward:1993mw}.

\item Inability of asymptotically predictable black holes to bifurcate \cite{Wald 1984}.

\item The domain of outer communication of a stationary, asymptotically flat spacetime is simply connected, if the domain is globally hyperbolic \cite{Chrusciel:1994tr} \cite{Galloway 1995}.

\item A stationary, asymptotically flat black hole has topology $\mathcal{S}^2$, if the domain of outer communication is globally hyperbolic and the closure of the black hole is compact  \cite{Galloway 1993} \cite{Chrusciel:1994tr}.

\item Most formulations of the black hole ``no Hair" theorem for asymptotically flat black holes
\cite{Israel 1967}, \cite{Israel 1968}, \cite{Carter:1971zc}, \cite{Bekenstein:1971hc}, \cite{Teitelboim 1972a}, \cite{Teitelboim 1972b}, \cite{Wald 1972}, \cite{Wald 1973}, \cite{Muller}, \cite{Robinson:1975bv}, \cite{Robinson 77}, \cite{Mazur}.

\item Generalized black holes are regions of ``no escape" \cite{Hayward:1994pj}.

\item Limits on energy extraction by gravitational radiation from colliding asymptotically flat black holes \cite{Hawking 1971}.

\item The Generalized Second Law of Thermodynamics \cite{Flanagan:1999jp}.

\item Bousso's covariant entropy bound \cite{Flanagan:1999jp}.

\item The  Shapiro ``time-delay" is always a delay, never an advance \cite{Visser:1998ua}.

\item Classical  Chronology  Protection  Conjecture \cite{Hawking 92}.

\end{enumerate}

\newpage

\section{Wormholes}
\label{WH}

There is a vast amount of literature on wormholes, from science fiction books and comics to rigorous mathematical papers. In principle, the existence of these objects does not lead to problems in physics, nor to strange time paradoxes. That is because in most cases wormholes are not \textit{traversable}. For example, a specific class of non-traversable wormholes the are the \textit{Einstein Rosen} (ER) bridges, which can be found in the simplest of spacetimes (sec. \ref{WH_Trav}). However, when wormholes can be traversed, even by a single light signal, things start to get out of hand. Causality violations, time travel and all the paradoxes that come with them, escape the Pandora's Box that is traversable wormholes, threatening to overthrow our current physical understanding of nature.     

\subsection{Wormhole Traversability}
\label{WH_Trav}

The first people to our knowledge to rigorously study \textit{traversable wormhole} spacetimes, were Morris and Thorne \cite{Morris:1988cz}. In their work, they start by discussing
several already existing solutions to the Einstein field equations that contained wormholes and argue against their use for interstellar travel. 
Furthermore, they discuss the type of properties the solution must have so that a ``traveler'' can have a pleasant trip and introduced 
their own solutions, whose only caveat is the ``Exotic'' (NEC violating) stress energy tensor. Although it is known that classically, the existence of traversable wormholes is prevented by the NEC, the 
non-compliance of QFT to NEC allows for their solution to be realized in the semi-classical gravity regime. In this project, we are less ambitious than our predecessors, in the sense that we are not studying wormholes for the purposes of human interstellar travel. We settle with studying the simple case of light. In other words, we investigate the conditions for super-luminal messaging through the wormhole. Hereon, we follow the structure of their discussion, reformulated in a way that fits the needs of our project. The first step, is to define what we mean by a traversable wormhole:

\begin{mydef}
\label{traversibility cond.} 
Traversable Wormhole: a Wormhole spacetime through which light signal can be sent back and forth.

\end{mydef}
 
\subsubsection{Schwarzschild Non-Traversable Wormhole}
\label{Schwarzschild Non-Traversable Wormhole}

First, we present a surprising example of a non-traversable wormhole. The wormhole which exists in the maximally extended Schwarzschild spacetime. Here, we briefly explain a few relevant to the discussion features of the said spacetime:

\begin{equation}
 ds^2 = -(1 - \frac{2G_N M}{r})dt^2 + (1 - \frac{2G_N M}{r})^{-1} dr^2  + r^2 (d \theta^2 + sin^2 \theta d\phi^2),    
\end{equation}
where $M$ is the mass of the Black Hole and $r$ is the radial distance from its center. This is the unique, spherically symmetric, static, vacuum solution
of the Einstein equations, which has a Horizon (outer-trapped surface) at $r=2G_N M $ and a singularity ($R^{\mu\nu\rho\sigma}R_{\mu\nu\rho\sigma}=\infty$) at $r=0$.
We  use the Kruskal–Szekeres coordinates, which have the advantage of covering the entire spacetime manifold of the maximally 
extended Schwarzschild solution, and are well-behaved everywhere outside the physical singularity $r=0$. In these coordinates, the metric takes 
the form:

\begin{equation}
 ds^2 =  \frac{32 G_N^3 M^3}{r}(-dT^2+dR^2)  + r^2 (d \theta^2 + sin^2 \theta d\phi^2) ,   
\end{equation}
where the new coordinates are:

\begin{equation}
 T = \pm (\frac{r}{2G_N M}-1)e^{\frac{r}{4G_N M}}\mbox{sinh}(\frac{t}{4G_N M}) ,   
\end{equation}

\begin{equation}
 R = \pm (\frac{r}{2G_N M}-1)e^{\frac{r}{4G_N M}}\mbox{cosh}(\frac{t}{4G_N M})  .  
\end{equation}

The minus sign describes the interior $0<r<2G_N M$, the plus the outside $2G_N M<r$ and the event horizon sits at
$T= \pm R$.

\bigskip

\begin{figure}[H]
\centering
\includegraphics[width=0.8\textwidth]{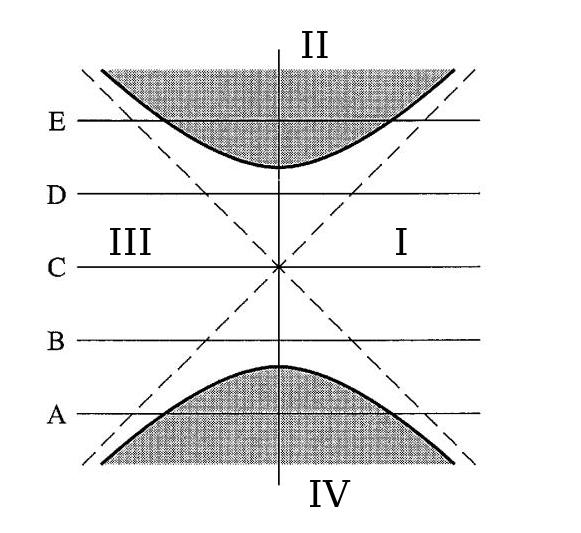}
\caption{ (Taken from \cite{Carroll:2004st})In these diagram T is the vertical and R the horizontal axis. Because the angular directions are suppressed, one should think of each point on this diagram as a 2-sphere of radius the value of coordinate $r$ at that point.}
\label{Kruskal}
\end{figure}

\bigskip

\begin{figure}[H]
\centering
\includegraphics[width=\textwidth]{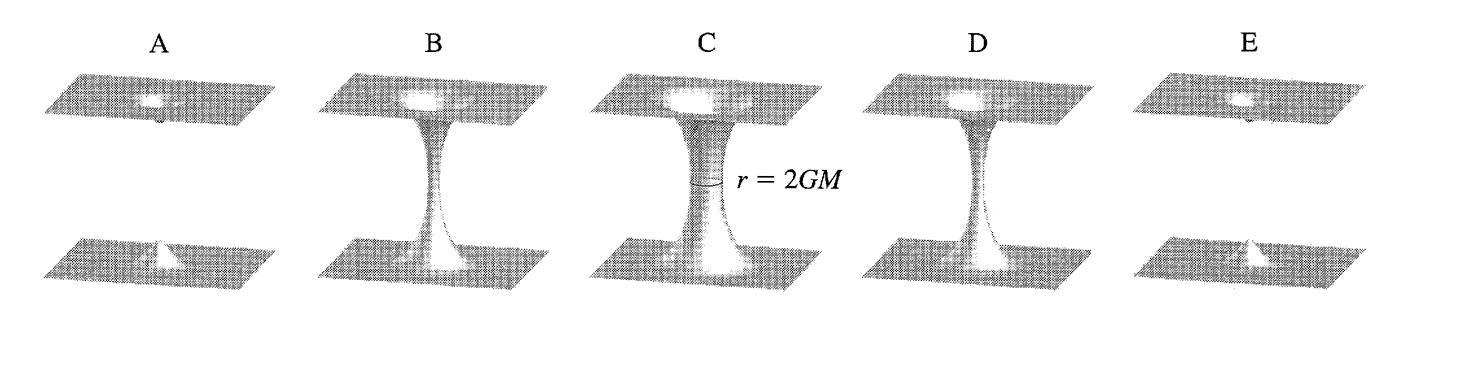}
\caption{(Taken from \cite{Carroll:2004st}) The spacelike slices of (fig. \ref{Kruskal}); Snap-shots of space, which show the formation and vanishing of the Schwarzschild dynamical wormhole.}
\label{slicing}
\end{figure}

To see the wormhole, we slice the Kruskal diagram (fig. \ref{Kruskal}) in \textit{spacelike} slices of constant $T$, and draw a picture of each slice (fig. \ref{slicing}) restoring
the angular coordinates. The pictures demonstrate the two asymptotically flat regions I and III, which are initially disconnected at the white hole's
anti-horizon, then they connect through a wormhole of maximal radius $2G_N M$, and finally become completely disconnected.

\bigskip

\bigskip

The first who noticed this ``dynamical wormhole'' of Schwarzschild spacetime, was Ludwig Flamm \cite{Flamm}, only a year after Einstein formulated his field equations. Although it
is intriguing that a wormhole shows up already in one of the plainest and well-studied GR's solutions, the connection 
between spacetimes I and III is spacelike. Thus, there is neither a null nor a timelike path that connects them, which means that the Schwarzschild wormhole is not traversable. 

\subsubsection{Ellis Traversable Wormhole}
\label{Ellis Traversable Wormhole}

Although we discussed only the Schwarzschild case, there are many more examples of spacetimes which contain non-traversable wormholes. Thus, it is more prudent to introduce a list of desired properties the wormhole spacetime \textit{should} have, to be considered traversable. Following \cite{Morris:1988cz}, we add some auxiliary properties that make the solution sufficiently easy to handle:

\bigskip

1) The spacetime metric should be \textit{spherically symmetric} and \textit{static}. This is purely for practical reason since 
having this amount of symmetry will greatly simplify our calculations.
\bigskip

2) The solution must obey \textit{everywhere} the Einstein equations
\bigskip

3) The solution must have a \textit{throat} that connects two \textit{asymptotically flat} regions of spacetime. Asymptotic flatness can, and has been relaxed, in several solutions of this form \cite{Krasnikov:2000gg}. 
\bigskip

4) There should be no \textit{horizon}, since it will prevent a two-way communication.
\bigskip

5) The solution should be perturbatively stable, meaning that it will not collapse when our photons try to cross its throat. The 
stability the wormholes defined here and more, has been studied in \cite{Hyunjoo Lee}, \cite{ArmendarizPicon:2002km}, \cite{Kuhfittig:2014bxa}.

\bigskip

6) The photons must be able to go through the wormhole's throat and back, in a reasonable amount of time. More specifically, the proper time it takes for the signal to reach an observer, measured by that observer, must be small.

\bigskip

7) The curvature should be low, compared to the \textit{Planck scale}. High curvatures, as we discuss in sec. \ref{VSCG}, signal the breakdown of the semiclassical gravity approximation. If that happens, one needs to consider \textit{Quantum Gravity} effects which would push the discussion beyond the scope of this work.  

\bigskip

8) There must be no singularity in our solution. Although this might seem a condition similar to the above, finiteness of the curvature scalars everywhere is only part of this restriction, as discussed in def. \ref{sing def}.

\bigskip

Now that we have our criteria, we are ready to write down a metric that satisfies them. To our knowledge, the following metric is 
the simplest traversable wormhole metric one can consider: 

\begin{equation}
\label{Ellis metric}
 ds^2 = -dt^2 +  dl^2  + (b^2 _0 + l^2) (d \theta^2 + sin^2 \theta d\phi^2) .   
\end{equation}

This solution's name is \textit{Ellis Wormhole}, and it is a special solution of the Ellis drain-hole \cite{Ellis}. The coordinates 
have the following ranges: $-\infty< t < +\infty$, $-\infty< l < +\infty$, $0\leq \theta \leq \pi$,  $0\leq \phi \leq 2\pi$ and $b_0$ is 
a constant. $t$ is the proper time a static observer measures, $\theta,\phi$ are the angular spherical coordinates, and $l$ is the 
proper radial distance at a fixed $t$. As required the spacetime is static, spherically symmetric and does not have horizons or 
singularities. The asymptotically flat regions lie at $l= \infty$ and $l= - \infty$.

\begin{figure}[H]
\centering
\includegraphics[width=0.65\textwidth]{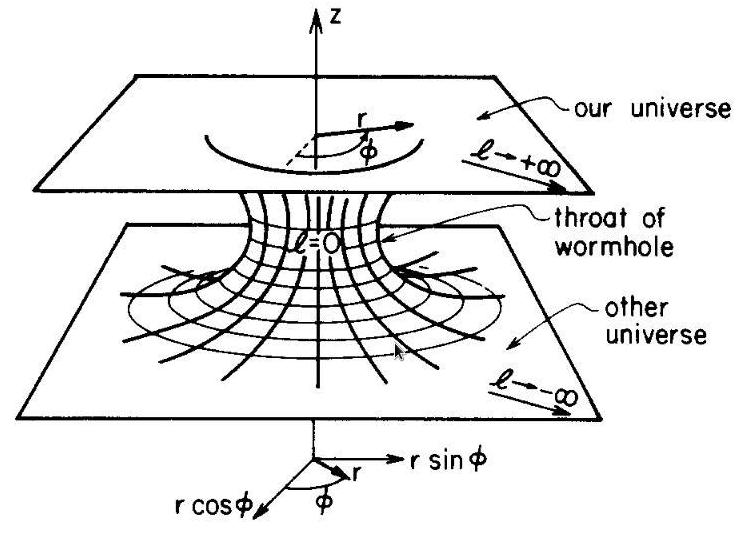}
\caption{(Taken from \cite{Morris:1988cz}) A wormhole connecting two asymptotically flat regions of spacetime. The diagram depicts this geometry for constant time $t=t_0$ and $\theta= \pi/2$ }
\label{Wormhole}
\end{figure}

If one wishes to obtain some visual confirmation of the wormhole in this spacetime, he/she can fix time and the $\theta$ angular
coordinate and 
embed
this 2-D spacelike slice in a 3-D Euclidean space. The resulting diagram is shown in fig. \ref{Wormhole}.  

%++subscript: use cylindrical co-ords bla bla, find in Thorne paper box 1 ++

\bigskip    

Finally, we study the most relevant quantity to this work; the stress energy tensor which corresponds to this spacetime.
Since we have the metric of the manifold, its computation is a rather trivial matter. We first compute the Einstein tensor $G_{\mu\nu}$, and then use the Einstein equations to obtain $T_{\mu\nu}$.
The components of the Einstein tensor after the usual simple but long calculation, are the following:

\begin{equation}
\begin{split}
 G_{tt} & =-\frac{b^2_0}{(b^2_0+l^2)^2}\;, \\
 G_{ll} & =-\frac{b^2_0}{(b^2_0+l^2)^2}\;, \\
 G_{\theta\theta} & =\frac{b^2_0}{(b^2_0+l^2)}\;, \\
 G_{\phi\phi} & =\frac{b^2_0 sin^2 \theta}{(b^2_{0}+l^2)}\;.
\end{split}
\end{equation}

Notice, that the corresponding $T_{\mu\nu}$ has some strange properties. Namely:
\begin{equation}
\begin{split}
T_{tt} & =-4\pi G_N \frac{b^2_0}{(b^2_0+l^2)^2}\;, \\
T_{ll} & =- 4\pi G_N\frac{b^2_0}{(b^2_0+l^2)^2}\;,
\end{split}
\end{equation}
have negative signs. In sec. \ref{EC} we saw that NEC requires $T_{\mu\nu}K^{\mu}K^{\nu} \geq 0$, but one can see that for a $K^{\mu} = (1,1,0,0)$ (in the $l$ direction) $T_{\mu\nu}K^{\mu}K^{\nu} \leq 0$. We also mentioned that reasonable, physical classical matter satisfies this condition. In light of 
this unusual nature, this type of matter content was coined by Morris and Thorne ``exotic''. The authors of \cite{Morris:1988cz} cite one of their discussions with Don Page, where it is  
argued that any wormhole spacetime \textit{must} contain such exotic matter in order to be traversable. This co-dependence of exotic matter and traversability takes the form of a condition, which goes by the name \textit{flaring-out condition}. To honor the legacy of \cite{Morris:1988cz}, we keep this name, although our definition of the condition is slightly different.

%\subsection{Traveling Through The W/H}
%\label{Trav}
%In the middle of our 
%maniforld there is a hole (genus 1 maniford)
%travel on the hypersurface bla  

\bigskip

\subsection{The Null Flaring-out Condition}
\label{Flaring}

 To see how the traversability of wormhole spacetimes restricts $T_{\mu\nu}$, we use the Focusing Theorem (th. \ref{FocThe}). We follow the evolution of a null congruence through the wormhole geometry depicted in fig. \ref{Wormhole}. 
The light rays on this diagram travel in the $l$ direction of the co-dimension one surface which includes the throat. Light rays start their journey through the throat as ingoing, following a hypersurface on which the expansion scalar is negative. However, the geometry of the surface requires that the expansion at the throat
becomes zero, and then starts to increase; otherwise, the light rays would never escape. Hence, the light rays need to defocus, i.e. the expansion scalar needs to be positive\footnote{ Bear in mind that no caustics form on the throat, so the congruence has no reason to end.}.

\begin{mydef}
\textbf{Flaring-out Condition:} Any traversable wormhole spacetime requires an increase of the expansion scalar: $\frac{d\theta}{d\lambda} > 0 $.
\end{mydef}
 This behavior clearly violates the focusing theorem, which states that  
$\frac{d\theta}{d\lambda} \leq 0 $ should always hold. As a consequence, a violation of NEC is implied. To show this, we use \eqref{focusing equation} :

\begin{equation}
\label{Flaring expansion}
 \frac{d\theta}{d\lambda} = -\frac{1}{2}\theta^2 - \sigma_{\mu\nu} \sigma^{\mu\nu} - T_{\mu\nu} K^\mu K^\nu \leq 0.
\end{equation}

The evolution of the expansion can become positive if one of the above quantities becomes positive enough. Since the only
quantity that is not a square of a real number is $T_{\mu\nu} K^\mu K^\nu$, only this term is capable of changing the 
overall sign
of the r.h.s. of \eqref{Flaring expansion}.
But NEC tells us that $T_{\mu\nu} K^\mu K^\nu \geq 0$, hence the traversability of wormholes requires the existence of NEC violating matter. This, as we saw, is true for our simple example of the Ellis wormhole, and for all traversable classical and semi-classical wormhole solutions we have seen in literature. 

 As we already
mentioned, this exotic $T_{\mu\nu}$ can be generated by a number of otherwise well-behaved states in QFT. Only, in this case, the classical Einstein equations cannot be used anymore, but a semiclassical version of them, in which the l.h.s. is the usual, fixed spacetime geometry described by $G_{\mu\nu}$ and the r.h.s is a quantum matter source given by $\langle T_{\mu\nu} \rangle$.

%write the computation from the Raychaudhuri paper review that computes the restriction of for W/H using NEC

\newpage

\section{Violations of NEC in QFT}
\label{Viol}

In this chapter, we discuss violations of the NEC in QFT. As we have already mentioned, if one wants to study such violations,
the first problem he/she encounters is to define a version of NEC in the context of QFT 
(sec. \ref{sec:The form of NEC in QFT}). Thereafter, we study some explicit 
examples in which these violations occur (sec. \ref{sec:1+1 CFT}, \ref{0+2}, \ref{Casimir}), and discuss their implications. The acute reader might have already guessed what fundamental feature of QFT leads 
to her rebellious behavior; nonetheless, we leave this discussion for later.

\subsection{The form of NEC in QFT}
\label{sec:The form of NEC in QFT}
 
  First, one should notice that the $T_{\mu\nu}$ that appears in a QFT, is not the same object that appears in the classical Einstein equations. In quantum theories, the 
fundamental objects that carry the information about the state of our system are wave-functions/states. The way to extract information about any observable
is to act on these states with the corresponding operators.

 In QFT, things are a bit more subtle, but the main idea is the same: if we want $T_{\mu\nu}$ to make sense, we need to promote it to an operator, while if we want to extract the observable quantities it describes, we need to act with it on a physical state. More explicitly, when the operator acts on a state which is its \textit{eigenstate}, the result is the eigenstate times its \textit{eigenvalue}, which we then read off as the value of the observable. So what happens if the state we consider is \textit{not} an eigenstate of the operator we are interested in? Any state can be decomposed in the basis of eigenvectors of our operator. The action of the operator will give us the sum of these superposed eigenstates, each multiplied by its energy eigenvalue times a weight (c-number). This should come as no surprise to us, since it is a direct manifestation of the quantum nature of the theory.
 
 Now, if one is determined to extract a single value, he can take the expectation value of the operator in this state. This is essentially the above sum with the following differences: a) the squared weights appear and they correspond to the probability with which each eigenvalue occurs, b) since the basis of eigenvectors is \textit{orthonormal}, cross-terms vanish and diagonal terms give the identity, leaving us with a sum of pure numbers.
 
 So every expectation value is an implicit sum of all possible eigenvalues of the operator, each multiplied by the appropriate probability. Clearly, it is uncertain if one should trust this ``average'' as the value of the observable he wanted to compute. An example in which one could imagine that this quantity makes sense, is a state for which the \textit{probability distribution} of the eigenvalues is sharply peaked around a single value. Then, one would have good reason to believe that the leading contribution is indeed given by that eigenvalue of the operator\footnote{In later chapter, we will see that not only the sharpness but the overall form of the probability distribution function matters. For example for a function with a long tail with no cut-off, a big contribution to the expectation value may come from some very big (compared to the mean) eigenvalue at the end of the tail (heavy tail).}.
 
 An obvious counter-example is a state in which the probability is evenly distributed among two eigenvalues, and thus the expectation value is not close to an actual eigenvalue of the operator, but some intermediate value. We shall see that the distinction between the two cases is connected to the \textit{fluctuation} of the operator we are interested in, and more accurately to its variance. The phenomenon which gives rise to this ``in between'' expectation value is \textit{quantum interference}.  This phenomenon is a central feature of quantum theories and deserves a more detailed discussion to which we turn to in the chapter of semiclassical gravity (ch. \ref{Semiclassical}). For now, let us define the form of a naive \textit{Quantum Null Energy Condition}, whose violations we will discuss in this chapter:

\begin{equation}
\label{naive QNEC}
 \langle \psi \rvert T_{\mu\nu} \lvert \psi \rangle K^\mu K^\nu \geq 0.
\end{equation}

The quantities involved in the above equation 
are the expectation value of $\hat{T}_{\mu\nu}$ for some state $\psi$ and a null vector $K^\mu$. 
It is a well-known fact that the expectation value of $\hat{T}_{\mu\nu}$ in QFT is a divergent, state-dependent object. Thus, one should regulate $\hat{T}_{\mu\nu}$ before computing $\langle \psi \rvert T_{\mu\nu} \lvert \psi \rangle$ and contracting it with the null vector.

\subsection{1+1 CFT}
\label{sec:1+1 CFT}

%A Conformal Field Theory in two dimension enjoys a huge amount of symmetry. 
% bla

For the following discussion, we suggest that the reader unfamiliar with \textit{Conformal Field Theories}(CFT's) takes a look at the introductory \cite{Tong:2009np} notes or/and the book \cite{Dif}.
We will introduce the tools that we shall use in our calculations shortly, but first some linguistics. By the word operator in CFT,
we mean a one or more fields, e.g. $\phi$ and $\phi \phi$, or some more complicated expression like $T_{\mu\nu}$. Moreover, since there is 
an operator-state correspondence, 
we may even refer to a whole state
as a field and vice versa. The reader will also encounter the terms \textit{primary} and \textit{first descendant}.
For the purposes of this chapter, he/she should think of them as nothing more than a ground state and the first excited state of the theory. 
1+1 CFT gives as the remarkable technology to construct a theory (Lagrangian) independent expression for the $\hat{T}_{\mu\nu}$ in terms of the Virasoro Generators, i.e. the generators of the conformal algebra.

Although the calculation we perform here is simple, the author of this thesis does not know of any piece of literature which includes it.
 We have chosen to do our calculations in the flat 1+1-D case for a non-interacting CFT since the
 message we are trying to convey does not require further 
complexification. Nevertheless, it is possible to generalize our results to higher dimensional CFT's, and curved backgrounds. 

We start by picking a state that we believe will exhibit (Q)NEC violating behavior. This state is a superposition of a primary field $O_{2}\lvert 0 \rangle$ 
of conformal weight $h_2$ and its first descendant  $O_{1}\lvert 0 \rangle$. As we already discussed, the descendant of a primary field/state is created 
by the action: $O_{1}\lvert 0 \rangle = L_{-1} O_{2}\lvert 0 \rangle$. The route we choose for our calculation is the following:

1) We compute the expectation value of the holomorphic  part of $\hat{T}_{\mu\nu}$ on the Euclidean plane.
\medskip

2) We conformally transform it to a Euclidean cylinder.

\medskip
3) We Wick rotate to the
Lorentzian signature,
to obtain the final expression for $\hat{T}_{\mu\nu}$.
\medskip

%* Include full computation, write it in the appendix, or nowhere? * 

As we mentioned, the holomorphic and anti-holomorphic part of $T_{\mu\nu}$ on the circle can be written in terms of Virasoro generators, a fact which greatly simplifies our calculations. The expressions are:

\begin{equation}
 T_{zz}= \sum_{n=-\infty}^{\infty} L_n z^{-n-2}\;,
\end{equation}

 \begin{equation}
 \widetilde{T}_{\widetilde{z}\widetilde{z}}= \sum_{n=-\infty}^{\infty} \widetilde{L}_n \widetilde{z}^{-n-2}\;.
\end{equation}
where the terms with tildes in the expression are the anti-holomorphic ones. For the computation we also need the Virasoro commutation relations:

\begin{equation}
 [L_{m},L_{n}]= (m-n)L_{m+n} + \frac{c}{12}(m^{3}-m)\delta_{m+n,0}\;.
\end{equation}

The state $O_{2}\lvert 0 \rangle = \lvert O_{2} \rangle $ is normalized, i.e.
 $\langle O_{2} \lvert O_{2} \rangle =1 $, but the first excited state $\langle O_{1} \lvert O_{1} \rangle = 2h_{2} $ is not. We choose to normalize the overall
superposed state in the following way: 

 \begin{equation}
  N^2 (\langle O_{1} \rvert - \langle O_{2} \rvert)( \lvert O_{1} \rangle - \lvert O_{2} \rangle) =1 \implies N= \frac{1}{(h_2+1)^{1/2}}\;.   
 \end{equation}

We should note here that although we have chosen to define the superposed state with a minus relative sign between the primary and the descendant, a positive relative sign would give an almost identical result. We can now
proceed with our computation of the $\langle \psi \rvert T_{zz} \vert \psi \rangle $:

\begin{equation}
\label{SET_holo}
 (\langle O_{1} \rvert - \langle O_{2} \rvert)T_{zz}( \lvert O_{1} \rangle - \lvert O_{2} \rangle)= N^2 \left[\left(3h_2 + 2h^2_{2} \right)z^{-2}-2h_2 z^{-1}-2h_2 z^{-3}
\right]\;,
\end{equation}

\begin{equation}
 \left(\langle O_{1} \rvert - \langle O_{2} \rvert \right)\widetilde{T}_{\widetilde{z}\widetilde{z}} \left( \lvert O_{1} \rangle - \lvert O_{2} \rangle \right)=
 N^2 \left[\left(3\widetilde{h}_2 + 2\widetilde{h}^2_{2} \right)\widetilde{z}^{-2}-2\widetilde{h}_2 \widetilde{z}^{-1}-2\widetilde{h}_2 \widetilde{z}^{-3}\right]\;.
\end{equation}

Notice the generality of this result, a common feature of calculations in 1+1 CFT. We have not chosen a specific conformal weight yet,
 thus the above result holds for \textit{any} primary. All one has
 to do is to pick
 his/her favorite theory  
plug in the conformal weight of his/her chosen primary and obtain $\langle O_{1} \rvert - \langle O_{2} \rvert)T_{zz}( \lvert O_{1} \rangle - \lvert O_{2} \rangle)$. However,
we do not stop our calculation here, since we would like the 
quantities described by the stress-energy tensor to be physical observables and in Euclidean coordinates that is not always the case. To go from the stress-energy tensor on the Euclidean plane to the one on the conformally equivalent Euclidean cylinder, one has to perform a 
conformal transformation. The stress-energy tensor transforms in the following way: 

\begin{equation}
 \langle T'_{z'z'}(z') \rangle= \left( \frac{\partial z'}{\partial z}\right)^{-2} \left[T_{zz}(z)-\frac{c}{12}S(z',z)\right] .
\end{equation}

Where $z'$ is the cylinder and $z$ the plane coordinates and $S(z',z)$ is the \textit{Schwartzian} derivative. %*write computation?*.
We are going to
compute only the holomorphic part, since the calculation for the anti-holomorphic part is exactly the same. After computing the Schwartzian and the other derivatives, 
we obtain:

\begin{equation}
 \langle T_{cyl}(z') \rangle=-z^2 \langle T_{plane}(z) \rangle + \left\langle \frac{c}{24} \right\rangle .
\end{equation}

Plugging in the $T_{zz}$ from \eqref{SET_holo}, we get the holomorphic part of the stress-energy tensor on the cylinder:

\begin{equation}
 \langle T_{z'z'} \rangle =N^2 \left[\left(3h_2 + 2h^2_{2} \right)-2h_2 z-2h_2 z^{-1}\right] + \left\langle \frac{c}{24} \right\rangle .
\end{equation}

To obtain the desired Lorentzian component $T_{\tau\tau}$, we perform a Wick rotation:

\begin{equation} \label{SET_Lor}
\begin{split}
 \langle T_{\tau\tau} \rangle & = - \langle T_{z'z'} \rangle- \langle \widetilde T_{\widetilde z' \widetilde z'} \rangle \\ & = N^2 \Big[ \left(3h_2 + 2h^2_{2}
 + \mbox{anti.-hol.} \right) -\left(4h_2 \cos(\sigma-\tau)+ \mbox{anti.-hol.}\right) - \\& - \left\langle \frac{c}{24} \right\rangle -
\left\langle \frac{c}{24}2h_2 \right\rangle \Big] .
\end{split}
\end{equation}

Similarly, we can  
compute the $\langle T_{\sigma\sigma} \rangle$ component. % write it
Our result \eqref{SET_Lor} allows us to calculate the expectation value of the energy momentum tensor for the superposed state
 $N (\langle O_{1} \rvert - \langle O_{2} \rvert)$  in \textit{any} 1+1 CFT for \textit{any} primary of conformal weight $h_2$.
It is easy to see that the maximum negative value of the $\langle T_{\tau\tau} \rangle$ is reached when we are on the light-cone of Lorentzian cylinder.
For that value we plot $\langle T_{++} \rangle$ against 
$h_2$, in other words, we see the flow of the expectation value of the SET in the \textit{space of primaries}:

\begin{figure}[H]
\centering
\includegraphics[width=0.8\textwidth]{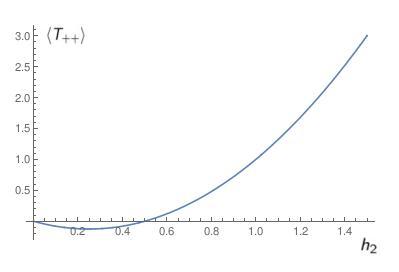}
\caption{$\langle T_{++} \rangle$ plotted against the conformal weight of $\mathcal{O}_2$.}
\label{h2}
\end{figure}

Another interesting plot is the one that describes how the $\langle T_{++} \rangle$ component of the stress tensor evolves with $\sigma$ and $\tau$
for a specific primary. Our choice is the vertex operator of the free scalar theory, whose conformal weight is $h_2 \sim k^2$ where $k$ is the momentum. From the above we see that for $h_2 \geq 0.5$, the state is not NEC violating, so we choose the intermediate value $h_2 = 0.2$:

\begin{figure}[H]
\centering
\includegraphics[width=0.8\textwidth]{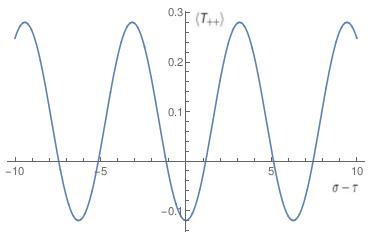}
\caption{$\langle T_{++} \rangle$, when the conformal weight of $\mathcal{O}_2$ is  $h_{2} = 0.2$, plotted against $\sigma-\tau$ ($x_-$ \mbox{in light-cone coordinates}). }
\label{cos}
\end{figure}

\bigskip

\textit{\textbf{Comment}}: An important feature of our result is that the negativity of the expectation value emerges from an 
oscillatory term. If we wish to compute the \textit{overall} energy, we should integrate this quantity over the whole space, which is the 
general procedure one follows to obtain a conserved charge (Energy) from its conserved current ($T_{\mu\nu}$). The integration over the oscillatory term
will clearly yield zero, hence the result we will obtain will be \textit{strictly} positive. The \textit{locality} of the effect is a hint that the NEC violating behavior of these states is connected to interference.

\bigskip

\subsection{Vacuum plus 2 particles}
\label{0+2}
In this section, we turn to the free massless scalar field theory on 4-D Minkowski spacetime. The state considered is a superposition of the vacuum plus two modes 
with \textit{specified} momentum k. The calculation of the expectation value of the stress tensor in this state was first produced by L.H.Ford and M.J. Pfenning \cite{Pfenning:1998ua}. A calculation we followed, with some slight differences that we point out along the way.

\subsubsection{Framework}
We start by performing the usual quantization procedure. The action is:

\begin{equation}
 S=\int_{-\infty}^{\infty} (-g)^{1/2} d^4x \left(\frac{1}{2}\partial_\lambda \phi \partial^\lambda \phi \right)\;.
\end{equation}

It is customary to define the stress-energy tensor in this procedure as:

\begin{equation}
 T_{\mu\nu}= -2 \frac{\delta \mathcal{S}_{matter}}{ \sqrt[]{-g}~ \delta g_{\mu\nu}} \;.
\end{equation}

Thus, for the Minkowski spacetime we get:

\begin{equation}
\label{SET}
 T_{\mu\nu}= \partial_\mu \phi \partial_\nu \phi - \frac{1}{2} \eta_{\mu\nu} \partial_\lambda \phi \partial^\lambda \phi\;.
\end{equation}

One difference between the Ford-Pfenning \cite{Pfenning:1998ua}
calculation and ours is that we will not consider the part proportional to the metric. This is simply because 
at the end of the day we will contract the expectation value of the full stress tensor twice with a null vector, thus the second term on the r.h.s. of \eqref{SET} will vanish. 

The scalar $\phi(x)$ we considered is a function of spacetime and must be a solution to the classical Equations Of Motion (EOM), obtained by variating the above action w.r.t. it. 
It is easy to see that the EOM is simply the Klein-Gordon equation:

\begin{equation}
 \partial_\mu \partial^\mu \phi(x) = 0.
\end{equation}
  
A solution to this equation is the plane wave, thus we Fourier expand $\phi(x)$ in terms of a complete set of plane wave mode function:

\begin{equation}
\label{mode functions}
 f_k (x)= \left(2wL^3 \right)^{-(1/2)} e^{ i \textbf{k x }-iwt}.
\end{equation}

By $x$ we denote spacetime coordinates, by bold $\textbf{k}, \textbf{x}$ space coordinates and by $t$ the usual Minkowski time. 
The Fourier expansion of $\phi(x)$ in terms of annihilation-creation operators and mode functions is:

\begin{equation}
 \phi(x)= \sum_{k}{} [a_k f_k (x) + a^\dagger_k f^*_k (x)  ].
\end{equation}

The next step is to require that $\hat{a},\hat{a}^\dagger$'s obey the usual commutation relations:

\begin{equation}
\label{com_a}
\begin{split}
[\hat{a}_k,\hat{a}^{\dagger}_{k'}] & = \delta_{kk'}, \\
[\hat{a}_k,\hat{a}_{k'}] & = 0, \\
[\hat{a}^{\dagger}_k,\hat{a}^{\dagger}_{k'}] & = 0.
\end{split} 
\end{equation}                               

We also define the vacuum as the state annihilated by annihilation operators $\hat{a}_k$ and the first excited state as the state created by 
the action of one $\hat{a}^{\dagger}_k$ on the vacuum. In formulas:

\begin{equation}
\begin{split}
 \hat{a}_k \lvert 0 \rangle &= 0, \\ \hat{a}^{\dagger}_k  \lvert 0 \rangle  & =  \lvert 1_k \rangle .
\end{split}
\end{equation}

It is also useful to define the action of $\hat{a},\hat{a}^\dagger$'s on a state of the Fock space that includes $n$ modes with $k_{th}$ momentum, i.e.:

\begin{equation}
\begin{split}
 \hat{a}^{\dagger}_k \lvert n_k \rangle &= \sqrt{n+1} \lvert {n+1}_k \rangle, \\ \hat{a}_k  \lvert n_k \rangle  & = \sqrt{n} \lvert {n-1}_k  \rangle.
\end{split}
\end{equation}

 The field $\phi$ gets promoted to an operator $\hat{\phi}$ with again the usual commutation relations and we have thus quantized our theory. All that is left to do, is plug $\hat{\phi}$ in the formula for the stress tensor \eqref{SET}, which is then also promoted to an operator, and 
compute its expectation value in the state of interest. We will compute the expectation value of each component of $\hat{T}_{\mu\nu}$ separately so that 
the negative energy density contribution becomes manifest from the start. From now on we drop the hats for simplicity, bearing in mind that all $T_{\mu\nu}$ and $a_k,a^{\dagger}_k$'s are operators:

\begin{equation}
\label{Ttt}
\begin{split}
 T_{tt} & =  \sum_{k}{}\sum_{k'}{}\left[(-iw) a_k f_k + (+iw) a^{\dagger}_k f^*_k \right] \left[(-iw') a_{k'} f_{k'} + (iw') a^{\dagger}_{k'} f^*_{k'} \right] \\
& =  \sum_{k}{}\sum_{k'}{} ww' \left[- a_k a_{k'} f_k f_{k'} + a_k a^{\dagger}_{k'} f_k f^*_{k'} + a^{\dagger}_{k} a_{k'} f_{k'} f^*_k 
 - a^{\dagger}_{k} a^{\dagger}_{k'} f^*_{k'}f^*_{k'} \right],
\end{split}
\end{equation}
where we have used the fact that $\partial_t \phi = -iw \phi$ and $\partial_t \phi^{\dagger} = iw \phi$. There is one more obstacle one needs to 
 surpass before obtaining the final result, and that is the divergent Vacuum Expectation Value (VEV) of the stress-energy tensor. That is: 

\begin{equation}
 \langle 0 \rvert T_{tt} \lvert 0 \rangle = \frac{1}{2L^3} \sum_{k}{} w = \infty.
\end{equation}
 
 In QFT, the set of techniques used to treat such infinities,
 is called \textit{renormalization}.
 The renormalization scheme we follow is \textit{normal ordering} and subtraction of the VEV of the stress energy tensor. To justify this
subtraction of the VEV one should remember that in QFT, we only measure energy difference,
and not the energies 
themselves. Thus, it is common practice to compute the renormalized energy densities whose values are taken relative to the infinity background
energy. A quantity which involves annihilation and creation operators is said to be normal ordered when all creation operators are on the left
side of all annihilators. After normal ordering taking into account the commutation relations \eqref{com_a}, \eqref{Ttt} becomes:

\begin{equation}
\label{Ttt2}
\begin{split}
 T_{tt}  = & \sum_{k}{}\sum_{k'}{} ww' \left[- a_k a_{k'} f_k f_{k'} + a^{\dagger}_{k'} a_k  f_k f^*_{k'} + a^{\dagger}_{k} a_{k'} f_{k'} f^*_k 
 - a^{\dagger}_{k} a^{\dagger}_{k'} f^*_{k'}f^*_{k'} \right] + \\& + \frac{1}{2L^3} \sum_{k}{} w ,
\end{split}
\end{equation}
where the last term is exactly the divergent VEV. The fact that the vacuum term arises as the divergence of the $T_{tt}$ gives additional
justification to our renormalization scheme. Hence, the normal ordered finite energy density component of SET is given by the operator statement:

\begin{equation}
\label{no}
 : T_{tt} : = T_{tt} - \langle 0 \rvert T_{tt} \lvert 0 \rangle.
\end{equation}

\bigskip

Using the operator on the l.h.s. of  \eqref{no} we can obtain non-divergent expectation values for different physical states.

\subsubsection{Calculation}

As we mentioned above, a state that will give rise to a negative energy density is a certain superposition of the vacuum plus
two excited modes of the same momentum. For our calculations we
choose a normalization which gives the state the following form:

\begin{equation}
\label{0+2 state}
 \lvert \psi \rangle = \frac{\sqrt{3}}{2} \lvert 0_k \rangle + \frac{1}{2} \lvert 2_k \rangle .
\end{equation}

Now we are ready to perform the calculation of the expectation value of the operator $: T_{tt} :$ in the above state. Using \eqref{no} and 
\eqref{Ttt2}:

\begin{equation}
\langle \psi \rvert : T_{tt} : \lvert \psi \rangle =  \frac{\sqrt{3}}{4} \langle 0_k \rvert : T_{tt} : \lvert 2_k \rangle +  \frac{\sqrt{3}}{4} \langle 2_k \rvert : T_{tt} : \lvert 0_k \rangle
 +  \frac{1}{4} \langle 2_k \rvert : T_{tt} : \lvert 2_k \rangle.
\end{equation}
% * <isaias.roldan.alemany@gmail.com> 2016-07-01T10:21:59.292Z:
%
% Here you are ignoring the coefficients, right? It would be nice to say it or use a \sim symbol
%Although to be honest, it wouldn't surprise me that this is already something standard. 
%Or are there inside the states and that is why they appear in next equation
%
% ^.
The cross-terms after some trivial algebra yield:

\begin{equation}
\begin{split}
\frac{\sqrt{3}}{4} \langle 0_k \rvert : T_{tt} : \lvert 2_k \rangle + \frac{\sqrt{3}}{4} \langle 2_k \rvert : T_{tt} : \lvert 0_k \rangle & = w^2 \left(-\frac{\sqrt{6}}{4}\right) (f_k)^2 +
 w^2 \left(-\frac{\sqrt{6}}{4}\right) (f^*_{k})^2  \\ & = - \frac{w}{2L^3} \frac{\sqrt{6}}{2} \cos(\textbf{kx}-wt).
\end{split}
\end{equation}
% * <isaias.roldan.alemany@gmail.com> 2016-07-01T10:31:44.355Z:
%
% In oder to obtain this result in the previous equation you need the minus sign in the last term. (I suspect)
%
% ^.
And the diagonal term:

\begin{equation}
 \frac{1}{4} \langle 2_k \rvert : T_{tt} : \lvert 2_k \rangle = \frac{w}{2L^3}\;.
\end{equation}
% * <isaias.roldan.alemany@gmail.com> 2016-07-01T10:42:55.708Z:
%
% Here I am missing the coefficients again, but I would say that in your previous calculation you used them.
%
% ^.

Thus, we obtain the renormalized expectation value of the energy density of the vacuum plus two particles state in the massless free scalar theory:

\begin{equation}
\label{0+2 SET}
 \langle \psi \rvert : T_{tt} : \lvert \psi \rangle =  \frac{w}{2L^3}\left[1- \frac{\sqrt{6}}{2} \cos(\textbf{kx}-wt) \right].
\end{equation}
% * <isaias.roldan.alemany@gmail.com> 2016-07-01T10:44:22.716Z:
%
% Review this equation if you have corrected the previous ones
%
% ^.
 We notice again that the negative contribution comes from an oscillating term. 
% ?say some stuff about different normalizations?
%plot?? at least the cos plot

%The case when the energy can become infinitely negative!! + the case where this becomes a coherent state (LASER)

\subsection{Casimir}
\label{Casimir}

In 1948, the Dutch physicist Hendrik Brugt Gerhard Casimir showed that the vacuum energy density for the electromagnetic field between two perfectly conducting plates, is negative relative to the value of the Minkowski vacuum \cite{Casimir:1948dh}. A heuristic explanation is that certain modes are missing in the spacetime region between the plates. Namely, modes with a wavelength longer than the distance $L$ between the plates simply do not ``fit''. Therefore, compared to the Minkowski vacuum which contains all modes, the energy density is necessarily lower.

The counterparts of this electromagnetic effect have been computed by several authors \cite{Birrel}, \cite{DeWitt:1975ys} for the case of the scalar field. In this section, we quote their results for the cases of \textit{minimal} and \textit{conformal coupling} for the interaction between gravity and a massive scalar field, in 4-D.

The expectation value of the full stress energy tensor between the plates, for a conformal coupling to gravity:

\begin{equation}
\label{Casimir conf}
\langle 0 \lvert T_{\mu\nu} \rvert 0 \rangle_{conformal} = \frac{\pi^2}{1440 L^4} \begin{bmatrix}
       -1 & 0 & 0  & 0 \\[0.3em]
       0  & 1 & 0 & 0 \\[0.3em]
       0  & 0 & 1 & 0 \\[0.3em]
       0  & 0 & 0 & -3 \\[0.3em]
     \end{bmatrix},
\end{equation}
and for a minimal coupling:

\begin{equation}
\label{Casimir min.}
\langle 0 \lvert T_{\mu\nu} \rvert 0 \rangle_{minimal}= \langle 0 \lvert T_{\mu\nu} \rvert 0 \rangle_{conformal} + \frac{\pi^2}{48 L^4} \frac{(3-2\sin^2(\frac{\pi z}{L}))}{\sin^4(\frac{\pi z}{L})} \begin{bmatrix}
       -1 & 0 & 0  & 0        \\[0.3em]
       0 & 1           & 0 & 0 \\[0.3em]
       0           & 0 & 1 & 0 \\[0.3em]
       0           & 0 & 0 & 0 \\[0.3em]
     \end{bmatrix},
\end{equation}
where L is the distance between the plates in the z-axis. To explicitly see that these are indeed NEC violating stress tensors, one can pick a null vector, e.g:

\begin{equation}
K^{\mu}= \begin{pmatrix}
-1 \\
0 \\
0 \\
1
\end{pmatrix},
\end{equation}

%++ align the OCD inducing matrices ++ 

and compute the contaction of interest:

\begin{equation}
\begin{split}
\langle T_{\mu\nu} \rangle_{conformal} K^{\mu} K^{\nu} & = \langle T_{xx} \rangle_{conformal} + \langle T_{zz} \rangle_{conformal} \\
& =  - \frac{\pi^2}{360 L^4}.
\end{split}
\end{equation}

Similarly to the case of the Ellis wormhole \eqref{Ellis metric}, NEC is violated again in only one direction, namely z. Several authors, e.g. \cite{Butcher:2014lea}, have tried to find a stress tensor for the Casimir effect inside a wormhole. In this case, the effect arises due to the topology of spacetime itself, and more precisely the hole which contains the throat. The region in the neighborhood of the throat cannot contain modes with a diameter bigger than that of the throat. The $\langle\hat{T}_{\mu\nu} \rangle_{casimir}$, even for a very long and thin throat is zero for null rays directed exactly parallel to the throat. 

However, this should have been expected even from studying the simple Ellis wormhole. In sec. \ref{WH_Trav} we showed that the direction in which the violation should occur is the l direction, i.e. the direction \textit{parallel} to the ``walls'' which contain the throat. Here, we showed that the violation occurs in the direction \textit{normal} to the plates, which in the wormhole case is the direction vertical to the walls. Thus the NEC violating directions do not match.

Our simple suggestion, if one insists to use a Casimir stress tensor, is to place two conducting plates in the direction vertical to the ``walls''. This will give rise to a stress tensor, which at least violates NEC in the right direction. 

%put the above in future goals

%\subsection{Squeezed/Coherent states}
%\label{squeez}

\subsection{Perturbed vacuum states}
\label{perturbed}

An interesting, general class of states are the states which are close to the vacuum. A way to obtain these states is:

\begin{equation}
e^{i\int J(x)d^{d}x\Phi(x)}\lvert 0 \rangle,
\end{equation}
where $J(x)$ is a small source\footnote{$J(x)$ here is a real source, and not the purely mathematical entity one uses when he wants to obtain n-point functions from the path integral.}, and $\Phi(x)$ a free massless scalar. This a nice way to write such states, since one can expand in powers of $J$ and pick up more and more contributions to the expectation value of the stress tensor in these states.
%++ more accuracy++ .
One of the reasons why these states are interesting is that we the freedom to choose function $J$ to be any function we want.

Below, we compute the expectation value of the stress tensor for this state, in 1+1-D light-cone coordinates in Minkowski spacetime. However, our result can be generalized to higher dimensions. The stress tensor for the free massless scalar in light-cone coordinates has the components:

%introduce the light cone coordinates explicitly, like $z_+=x-t$ bla.. 

\begin{equation}
\begin{split}
T_{++} & = \partial_{+}\Phi(z)\partial_{+}\Phi(z), \\
T_{--} & = \partial_{-}\Phi(z)\partial_{-}\Phi(z).
\end{split}
\end{equation}

where we have chosen to work in the coincident limit. From now on, we use the notation $\Phi(z) = \Phi_z$ for aesthetic reasons. Next, we compute the expectation value of $T_{++}$ up to order $J^2$, using the expansion:

\begin{equation}
\label{T++ perturbed}
\begin{split}
\langle T_{++} \rangle_{per.} & = \langle 0 \rvert e^{-i\int J_x d^{2}x\Phi_x} T_{++} e^{i\int J_x d^2x\Phi_x}\lvert 0 \rangle \\
& = \langle 0 \rvert \left(1 -i\int d^2x J_{x}\Phi_{x}-\frac{1}{2}\int d^2x d^2w J_{x}J_{w}\Phi_{x}\Phi_{w} \right) \cdot \left( \partial_{+}\Phi_{z}\partial_{+}\Phi_{z} \right) \cdot \\ 
& ~~~~ \cdot \left(1 + i\int d^2w J_{w}\Phi_{w}-\frac{1}{2}\int d^2x d^2w J_{x}J_{w}\Phi_{x}\Phi_{w} \right)  \lvert 0 \rangle.
\end{split}
\end{equation}

\begin{enumerate}
\item Order $J^0$: one obtains the usual VEV, $\langle 0 \lvert T_{++} \rvert 0 \rangle$ which has already been extensively studied. % and is of no interest for this computation.

\item Order $J^1$: These are 3-point functions, and it is a well-known fact that for a free scalar field all odd n-point functions vanish. A heuristic way to understand this is by considering the Feynman diagram for a 3-point function; this is just the 3-point vertex. Since there are no interactions, there is no mechanism through which a ``particle'' can decay in two or two ``particles'' can combine in one. A more rigorous way to see this is to turn to the path integral formalism and obtain the n-point functions using the random source-functional derivative trick. He/she will then arrive at the same conclusion. 

\item Order $J^2$: These terms do contribute to \eqref{T++ perturbed} and we shall devote the rest of this section in calculating this contribution. 
\end{enumerate}

There are three terms of order $J^2$. Our first step is to deal with their cumbersome 4-point functions, by invoking the \textit{Wick's Theorem}:

\begin{equation}
\begin{split}
A & =  \langle   \Phi_{x}\partial_{+}\Phi_z \partial_{+}\Phi_z  \Phi_{w}   \rangle_0 \\
  & = 2 \cdot \langle   \Phi_{x} \partial_{+}\Phi_z \rangle_0 \cdot  \langle   \partial_{+}\Phi_z \Phi_{w} \rangle_0\;,
\end{split}
\end{equation}

\begin{equation}
\begin{split}
B & = - \frac{1}{2}\langle   \Phi_{x}\Phi_{w} \partial_{+}\Phi_z\partial_{+}\Phi_z \rangle_0 \\
  & = - \langle   \Phi_{x} \partial_{+}\Phi_z  \rangle_0 \cdot  \langle   \Phi_{w} \partial_{+}\Phi_z \rangle_0\;,
\end{split}
\end{equation}

\begin{equation}
\begin{split}
C & = - \frac{1}{2}\langle  \partial_{+}\Phi_z\partial_{+}\Phi_z \Phi_{x}\Phi_{w}  \rangle_0 \\
  & = - \langle \partial_{+}\Phi_z  \Phi_{x}   \rangle_0 \cdot  \langle  \partial_{+}\Phi_z \Phi_{w}  \rangle_0\;.
\end{split}
\end{equation}

By simply adding A,B and C one obtains:

\begin{equation}
\begin{split}
\label{A+B+C}
A+B+C & = \langle \partial_{+}\Phi_z  \Phi_{w}   \rangle_0 \cdot \left(   \langle   \Phi_{x} \partial_{+}\Phi_z \rangle_0 - \langle \partial_{+}\Phi_z  \Phi_{x}   \rangle_0 \right) + \\
& +  \langle \Phi_{x} \partial_{+}\Phi_z  \rangle_0 \cdot \left(   \langle \partial_{+}\Phi_z  \Phi_{w}   \rangle_0 - \langle   \Phi_{w} \partial_{+}\Phi_z \rangle_0  \right).
\end{split}
\end{equation}

Noticing that all operators are \textit{Hermitian}, using the identity $\langle AB \rangle =\langle BA \rangle^*$ and $\langle AB \rangle - \langle AB \rangle^*=2i \cdot \mbox{Im}\langle AB \rangle$ for \eqref{A+B+C} , we arrive to:

\begin{equation}
\begin{split}
\label{A+B+C 2}
A+B+C & = 2i \cdot \langle \partial_{+}\Phi_z  \Phi_{w}   \rangle_0 \cdot  \mbox{Im}  \langle   \Phi_{x} \partial_{+}\Phi_z \rangle_0  + \\
& + 2i \cdot  \langle \Phi_{x} \partial_{+}\Phi_z  \rangle_0 \cdot  \mbox{Im} \langle \partial_{+}\Phi_z  \Phi_{w}   \rangle_0 \;.
\end{split}
\end{equation}

The Feynman propagator in two dimensions reads: 

\begin{equation}
\label{Feynman prop}
\langle \Phi_{x} \Phi_z \rangle_0 = - \frac{1}{4\pi} \ln \frac{1}{(z_+ - x_+)(z_- -x_-)- i\epsilon } \;.
\end{equation}

% ++is there a minus in \ref{Feynman prop} ?? it doesn't rly matter but for completeness++

Remembering that all the derivatives are actually $\partial_+ = \frac{\partial}{\partial z^+}$, we use \eqref{Feynman prop} to compute all the relevant quantities in \eqref{A+B+C 2}:

\begin{equation}
  \langle   \Phi_{x} \partial_{+}\Phi_z \rangle_0  = \frac{\partial}{\partial z^+} \langle  \Phi_{x} \Phi_z \rangle_0 =  \frac{1}{4\pi}  \frac{1}{(z_+ - x_+)- i\epsilon }\;,
\end{equation}

\begin{equation}
\label{Im 1}
 \mbox{Im} \langle   \Phi_{x} \partial_{+}\Phi_z \rangle_0  =  \frac{\epsilon}{4\pi}  \frac{1}{(z_+ - x_+)^2- \epsilon^2 }\;,
\end{equation}

\begin{equation}
  \langle  \partial_{+}\Phi_z \Phi_w \rangle_0  = - \frac{\partial}{\partial z^+} \langle  \Phi_z \Phi_w \rangle_0 = - \frac{1}{4\pi} \frac{1}{(w_+ - z_+)- i\epsilon }\;,
\end{equation}

\begin{equation}
\label{Im 2}
 \mbox{Im}  \langle  \partial_{+}\Phi_z \Phi_w \rangle_0  =  \frac{\epsilon}{4\pi} \frac{1}{(w_+ - z_+)^2- \epsilon^2 }\;.
\end{equation}

% ++??should I say this?? The $\epsilon$ is just a mathematical artifact, introduced whenever analytical continuation of the divergent propagator is needed. Thus in the above calculation we have taken it out, computed the derivatives, and put it back in++ 

In the epsilon prescription we have used, there is always an implicit limit of $\epsilon \rightarrow 0$. Taking this limit we see that \eqref{Im 1} and \eqref{Im 2} are nothing more than an alternative definition for the delta function:

\begin{equation}
\label{delta func}
\delta(x)= \frac{1}{\pi} \lim_{\epsilon\to0} \frac{\epsilon}{x^2 + \epsilon^2}\;.
\end{equation}

After having brought the 4-point functions to the form we wanted, computing the integrals for the order $J^2$ terms becomes an easy task:

\begin{equation}
\label{Pert result}
\begin{split}
\mathcal{O}(J^2) : & -\frac{2i}{16\pi}\int dx_+ dx_- \int dw_+ dw_- J(x_+,x_-) J(w_+,w_-) \frac{1}{(w_+ - z_+)- i\epsilon } \delta (z_+-x_+) - \\
& -\frac{2i}{16\pi}\int dx_+ dx_- \int dw_+ dw_- J(x_+,x_-) J(w_+,w_-) \frac{1}{(z_+ - x_+)- i\epsilon } \delta (w_+-z_+) = \\
& = -\frac{2i}{16\pi}\int dx_- \int dw_+ dw_- J(c(x_+),x_-) J(w_+,w_-) \frac{1}{(w_+ - z_+)- i\epsilon } - \\
& -\frac{2i}{16\pi}\int dx_+ dx_- \int dw_- J(x_+,x_-) J(c'(w_+),w_-) \frac{1}{(z_+ - x_+)- i\epsilon } \;.
\end{split}
\end{equation}

% ++ be more explicit, find out wtf is going on with the sign ++  

Relabeling $x_- \rightarrow w_- $ the last to integrals combine, and we obtain:

\begin{equation}
\label{Pert result 2}
\begin{split}
\mathcal{O}(J^2) & =  -\frac{i}{8\pi}\int dx_+ dx_- \int dw_- J(x_+,x_-) J(z_+,w_-) \left( \frac{1}{(z_+ - x_+)- i\epsilon } - \frac{1}{(z_+ - x_+) + i\epsilon } \right) = \\ &  = -\frac{i}{8\pi}\int dx_+ dx_- \int dw_- J(x_+,x_-) J(z_+,w_-)  \frac{2i \epsilon }{(z_+ - x_+)^2 + \epsilon^2 }  = \\ &  = \frac{1}{4\pi}\int dx_+ dx_- \int dw_- J(x_+,x_-) J(z_+,w_-) \delta (x_+ - z_+) = \\ &  = \frac{1}{4\pi}   \left( \int dx_-  J(z_+,x_-)  \right)^2 \;.
\end{split}
\end{equation}

As we mentioned, J is a general function which describes the form of the source distribution. Finally, the result for the entire expectation value of the stress tensor up to second order in J is:

\begin{equation}
\langle T_{++} \rangle_{per.} = \langle T_{++} \rangle_0 + \frac{1}{4\pi}   \left( \int dx_-  J(z_+,x_-)  \right)^2.
\end{equation}

which hold for a general class of functions J. This result shows that there is a general class of states for which the quantity of interest is strictly positive. We take this as evidence that the spectrum of NEC violating states in QFT's is smaller than one would generally expect. 

\newpage

\section{Local and Non-Local Constraints}
\label{Constraints}

In the first three sections of this chapter, we present a selection of the most relevant constraints on exotic matter that currently exist in literature. We start by describing a local condition (sec. \ref{QNEC}), then discuss a global condition (sec. \ref{ANEC}) and finally turn to a ``semi-local'' one (sec. \ref{QI}). We have noticed that there are some general characteristics shared by members of each of the above categories. Local conditions, like NEC, are imposed at each and every point of spacetime. However, they lack restrictive power. The global ones are the most restrictive, but since they are imposed on an entire geodesic cannot be used to prove theorems for which point-wise conditions are necessary. Finally, the semi-classical conditions sit somewhere in the middle. They have good restrictive power, and are good candidates for proving more theorems than the global ones. However, they have their own issues, which we shall discuss at the end of sec. \ref{QI}.      
% Local bounds $\rightarrow$ not at all restrictive (divergence) of TT? 

% Global bounds $\rightarrow$ too vague, very restrictive

% Semi-classical bounds $\rightarrow$ more restrictive, not too vague

\subsection{QNEC}
\label{QNEC}

One of the most exciting, recent developments in the field of Energy Conditions is the \textit{Quantum Null Energy Condition}(QNEC), which was introduced by R.Bousso, A.Wall et al. \cite{Bousso:2015mna}. The goal of \cite{Bousso:2015mna} was to unify the \textit{Covariant Bousso bound} \cite{Bousso:1999xy}  with the classical focusing theorem (th. \ref{FocThe}), using a universal inequality, introduced by the same authors, called the \textit{Quantum Focusing Conjecture}(QFC). The QFC and how QNEC arises from it will be discussed in more detail in sec. \ref{QFC} .

An important feature of QNEC is that it is a statement independent of gravity; QNEC is a pure QFT statement. As such, it admits a proof within QFT, which was given by the same authors in \cite{Bousso:2015wca}. Another interesting property of QNEC is that, much like the classical NEC, it is a local condition, i.e. it is imposed on spacetime points. However, it does not hold for every point in spacetime for all spacetimes, but only for points that lie on stationary null surfaces\footnote{A null surface with expansion $\theta= 0$ everywhere on it.}. However, in Minkowski space, any point and null vector on it lie on such a surface, e.g a Rindler Horizon.  

Although QNEC has a clear meaning as an energy condition, the proof and even the interpretation of the quantities involved, is rather cumbersome. Hence, before we present the statement itself, we shall provide a detailed introduction, following closely the work in \cite{Bousso:2015wca}.

We start by picking a \textit{Cauchy Surface} $M$. A Cauchy surface is a spacelike slice, $t=constant$, which contains all information about our theory. Then, we place a QFT on this surface. Since the proof holds for spacetime dimensions $D \geq 2$, we work with a co-dimension one ($D-1$) surface. Next, we define a smooth co-dimension two surface $\Sigma$, which divides $M$ in two parts.

Since a QFT leaves on $M$, $\Sigma$ is now an entangling surface; the field theory degrees of freedom on one side of $M$ are entangled with the degrees of freedom which live on the other side. Hence, one can define $S_{out}$, which denotes the entanglement entropy of the fields residing in one side of $\Sigma$. Next we shoot a co-dimension one, \textit{Null Hupersurface} $N$, orthogonal to $\Sigma$. We pick a point $p$ which lies on $\Sigma$, and a null vector $k^{\mu}$. The vector $k^{\mu}$ is a member of a vector field $k^{\mu}(y)$ defined in the neighborhood of $p$, which is orthogonal to $\Sigma$ and tangent to $N$. Here, $y$ is a co-ordinate on $\Sigma$, called the ``transverse direction", which labels all possible $p$'s. 

The vector $k^{\mu} (p)$ generates a null geodesic on $N$, parametrized by an affine parameter $\lambda (p)$. Thus, the vector field which consists of all $k^{\mu} (y)$ in the neighborhood of $p$, defines a family of geodesics, parametrized by affine parameters $\lambda(y)$. By deforming $\Sigma$ along $N$ in the neighborhood of any $y$, one can obtain a family of surfaces which are now functionals of the affine parameter $\Sigma [\lambda(y)]$.

\begin{figure}[H]
\centering
\includegraphics[width=0.8\textwidth]{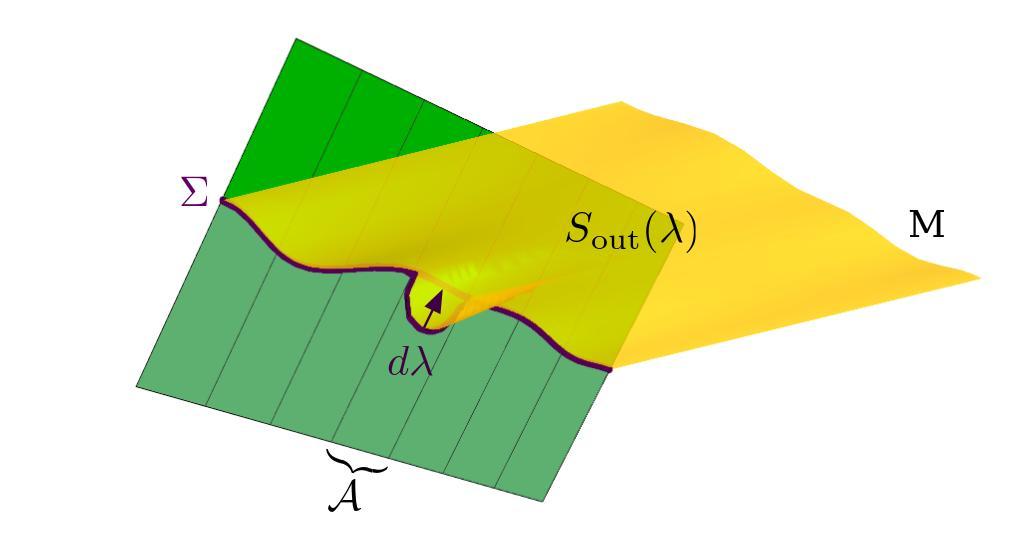}
\caption{ (Taken from \cite{Bousso:2015wca})Assuming the 4-D case and omitting one spatial dimension, one obtains the above figure. The null hypersurface $N$ is in green, in yellow we see the ``outside'' half of the Cauchy surface M and the black line is co-dim 2 surface $\Sigma$. The deformation of $\Sigma$, and hence the variation, is performed along a ``pencil'' of infinitesimal area $\mathcal{A}$, which contains point $p$. After deforming $\Sigma$ once more in the neighborhood of $p$, we obtain the second variation of the entanglement entropy of the fields living on half of M: $\frac{\delta^2 S_{out}}{\delta \lambda (p)\delta \lambda (p)}$.}
\label{QNEC figure}
\end{figure}

Finally, we can define a family of entanglement entropies $S_{out}[\lambda(y)]$ for all Cauchy splitting $\Sigma [\lambda(y)]$, which are the relevant quantities for the QNEC statement.

\begin{equation}
\langle T_{kk} (p)\rangle_\psi \geq \frac{\hbar}{2\pi \mathcal{A}} \frac{\delta^2 S_{out}}{\delta \lambda (p)\delta \lambda (p)} \;,
\end{equation}
 
where $T_{kk} (p)$ is the double contraction of the stress tensor of the QFT at point $p$ with the null vector $k^{\mu}$ normal to $\Sigma$ at $p$.
The r.h.s. is the diagonal term of the second functional derivative of $S_{out}[\lambda(y)]$. One of the reasons why QNEC is defined in such a complicated way is that $S_{out}[\lambda(y)]$ is not a local quantity. Thus, if one wants to involve such a quantity in strictly local bound like QNEC, he/she must find a way to distil the local information it contains. This is %+virtuosika+
done, by writing the entropy as a functional of $\lambda(y)$ and taking its \textit{derivative} at coincident points.

%\subsubsection{The state}

%Wall smears the field itself in order to do the null quantization. I have a problem with this since I believe it violates the one of the QFT Wightman axioms , cf.Reeds-Simon. Also, even if the smeared field is well-defined, the bound is in the end semi-local like QI... 

%Does the state have to be close to the Vacuum?? this is very troubling since in the Holographic proof of QNEC it is claimed that QNEC hold for \textit{any given state}!

%\subsubsection{The regime}
 
%1. Null quantization: valid Null-hypersurface initial-value formalism, existence of a field algebra $A(H)$ which can be defined on any stationary point on the horizon $H$ without making reference to anything outside $H$, all properties of algebra must be defined using no more than 1,2,3 

%2. Proven for $D \geq 2$  a) free super-renormalizable bosonic for any point that lies on a stationary null hypersurface, (Mink- Rindler bla bla) b) for theories with Einstein Gravity duals?? cf. Holographic proof of QNEC

%3. It is believed to hold for bla

% Look at Wall talk in KITP

\subsection{ANEC}
\label{ANEC}

Seeing our failure to generalize a \textit{local} NEC in QFT, an option is to consider \textit{non-local} conditions in hope that most, or at least some, of the results based on NEC can be salvaged. Historically the first to suggest such a condition was Tipler in 1977 \cite{Tipler:1978zz}. Tipler considered the average over one of these point-wise conditions, namely WEC (sec. \ref{EC}). The averaging was taken over the whole worldline of the observer, and the resulting bound is known as the \textit{Averaged Weak Energy Condition}. However, we are interested in NEC, so we shall discuss AWEC's counterpart, the \textit{Averaged Null Energy Condition}(ANEC):

\begin{equation}
\int^{\infty}_{-\infty} \langle T_{\mu\nu} K^\mu K^\nu \rangle d\lambda \geq 0 .
\end{equation}

The statement is the following: \textit{the integral over the entirety of a null geodesic with affine parameter $\lambda$, of the projected $T_{\mu\nu}$ on the null vector $K^{\mu}$ tangent to the geodesic, must always be greater or equal than zero}. A list of classical GR results that can be restored by imposing ANEC can be found in \cite{Curiel:2014zba}.  

An interesting aspect of ANEC is that it, much like QNEC, can be derived in QFT. In 1991, Wald and Yurtsever \cite{Wald:1991xn} proved that ANEC holds in two and four-dimensional Minkowski spacetime, for any \textit{Hadamard} state along an \textit{achronal} null geodesic. There are other proofs since then, like the one from Ford and Roman \cite{Ford:1994bj}, who proved the above using an entirely different method, or the one from Eleni-Alexandra Kontou and Ken D. Olum \cite{Kontou:2015bta}. In \cite{Kontou:2015bta} the above results are additionally generalized for spacetimes with small curvature. In a nutshell, ANEC holds for free fields in Minkowski spacetime but can be violated in some curved spacetimes. 

One may think that a traversable wormhole spacetime clearly violates ANEC, and if we accept ANEC as our new NEC for QFT such solutions should be considered unphysical. But there is a way out, even for wormholes as simple as the ones we considered in sec. \ref{WH_Trav} and the reason is that ANEC is defined as the integral over the ``whole'' null geodesic. Therefore, one could imagine placing the NEC violating matter in some local region of spacetime so that the flaring-out condition (sec. \ref{Flaring}) is satisfied and start throwing smaller and smaller pebbles of ``normal'' matter on his way to infinity. We would like our pebbles to be smaller and sparser as we reach infinity so that the asymptotic flatness condition is satisfied and our metric does not change drastically. This can be done in such a way, that the NEC violating energy the light ray ``sees'' is compensated by the NEC satisfying pebbles, resulting in geodesics that do not violate ANEC.      

\subsection{Quantum Inequalities}
\label{QI}

A set of semi-local conditions in QFT are the \textit{Quantum Inequalities}(QI's), originally introduced by Ford in 1978 \cite{Ford:1978qya}. The quantum inequalities are a bit more local than the averaged energy conditions since they involve integrals over only a specific interval of interest and not the whole spacetime. They are essentially restrictions on the magnitude/time duration on the negative energy density (or flux) in some time interval. The form of these restrictions is the following:

\begin{equation}
\langle \hat{\rho} \rangle =\int^{\infty}_{-\infty} \langle \rho \rangle f(\tau) d\tau  \geq  -\frac{C\hbar}{c^3 \tau^{4}_0}\;.
\end{equation}

Here $\rho$ is the renormalized expectation value of the energy density of a free field in an arbitrary Hadamard state, measured by an inertial observer in his rest frame. $\tau$ is the proper time measured by the observer and $f(\tau)$ is a sampling function (a smooth peaked function of width $\tau_0$, which integrates to 1). An example of such a function is the Lorentzian function:

\begin{equation}
f(\tau)= \frac{\tau_0}{\pi}\frac{1}{\tau^2 + \tau^2_0}\;.
\end{equation}

The above QI is the QI for the energy density and it states that the amount of the smeared negative energy density $\hat{\rho}$ is inversely proportional to the fourth power of the duration of the interval $\tau^4_0$. Simply put, the more negative energy you see, the shorter it can exist. It is interesting to notice that these bounds resemble the uncertainty principle, although it has not been assumed in their derivation.

%quantum interest? AWEC recovery in 4-D mink?

However, the sampling procedure does not come risk-free. If the probability distribution function of the eigenvalues of the operator that we sample is some strange function (not Gaussian, Lorentzian etc.) at least in the neighborhood of the expectation value, the sampled function may inherit features from the sampling function.

%Bad thing, their main result may be inherited behavior of the smearing function.

These constraints have been derived for free fields, namely: massless \& massive free scalar, E/M field, massive spin-1 field and Dirac field. They have been derived for \textit{arbitrary} smooth sampling functions in 2-D \& 4-D Minkowski and a number of curved spacetimes.  

Going back to our favorite traversable wormhole example we notice that from the bounds discussed so far, QI's place the strongest restriction in their existence. If one wishes to have a sufficiently big, macroscopic traversable wormhole so that he can travel through it comfortably, he would in principle need a huge amount of negative energy. This, in turn, implies that the wormhole would remain open for a ridiculously small amount of proper time. This is certainly bad news for sci-fi travelers. However, to their relief, there is still a way out. There exist wormhole solutions introduced by M.Visser et al. \cite{Visser:2003yf}, which require arbitrary small energy condition violations.

Furthermore, in this thesis we have defined a traversable wormhole as a wormhole spacetime through which a light signal can travel back and forth. Hence, for our purposes, neither the wormhole nor the period of its existence needs to be especially big. 

\subsubsection{ER=EPR and the Lost Wormhole}

One cannot help but notice, that an interesting question is raised in the above discussion. If the exotic matter can indeed only exist for a certain amount of time, what happens to the wormhole when this negative stress-energy vanishes? Clearly, the above question requires a detailed analysis of the dynamics of the problem. Instead, we give a heuristic description using intuitions coming from information theory and the ER=EPR proposal.

To justify our way of thinking we first discuss the dynamical Schwarzschild wormhole in the ER=EPR context. In the past few years, it has become a slogan of quantum information theorists that \textit{entanglement} is \textit{connectedness}. Following this logic, one should realize that although the proposal of \cite{Maldacena:2013xja} is somewhat ``fringy" at first sight, it has some reasonable intuitive origins. The proposal in its original form is the claim that pairs of maximally entangled black holes are connected via ER (Einstein-Rosen) bridges. In this picture, the entangled black holes are treated as a complex EPR (Einstein-Podolsky-Rosen) pair. For our purposes, a stronger version of the proposal also introduced in \cite{Maldacena:2013xja} is in order: every pair of entangled particles is connected via an ER bridge. Assuming the above, we proceed to a heuristic description of the interior of the Schwarzschild spacetime.

As we discussed in sec. \ref{WH_Trav}, following the evolution of spacelike slices (fig. \ref{slicing}) of the Kruskal diagram (fig. \ref{Kruskal}) one can see a non-traversable wormhole forming and then disappearing when the slices start ``touching'' the singularity. This means that for some period of time, the universes I and III are \textit{connected} and then are disconnected again. This connection signals the existence of entanglement between the degrees of freedom residing on universes I and III since entanglement is connectedness and vice versa. In the ER=EPR picture, things are even clearer. The existence of an ER bridge means that there are entangled pairs, whose one component leaves in I and the other in III.

Another factor which plays an important role to our description of this scenario is the breaking of the entanglement. According to the AMPS firewall proposal \cite{Almheiri:2012rt}, when entanglement is broken, a huge amount of energy is released. Regardless if one trusts the proposal or not, there is a simple lesson to take home. When entanglement is broken, energy is indeed released. To get some intuition as to why this is true, think about the following example. Take the vacuum state on a Cauchy slice with an entangling surface dividing it in two. Now assume that for some reason the entanglement between a single pair of field degrees of freedom is broken. Since the vacuum is \textit{defined} as the state of lower energy, the state you obtained will be by necessity of higher energy. 

Having acquired the required pieces, we present our picture starting from the point where axes T and R meet in fig. \ref{Kruskal}. As time passes, I and III become more and more disconnected and hence, more and more energy is released. When finally I and III are completely separate, a singularity has formed, i.e. the energy released from the breaking of the entanglement has become the mass of the black hole. Clearly, there is no simple observable that can allow one to test if this process takes place or not, since the interior is protected the whole time by the existence of the \textit{event horizon}. %However, it would be a fun calculation to compute the energy released by the entanglement and see if it indeed matches the mass of the black hole. %is that what the ADM mass is, where the entropy in this case is A/4G?

Next, we turn to our traversable wormhole case. It is clear from our definition (def. \ref{traversibility cond.}) that the solutions we are interested in cannot contain event horizons, so everything now is out in the open. In our discussion, the initial system is that of an Ellis wormhole, for simplicity. According to QI's the negative energy supporting the wormhole can exist for a certain amount of time. After that it gets annihilated via some, still not discussed, physical mechanism. More precisely, according to the \textit{Quantum Interest Conjecture} \cite{Ford:1999qv} , it gets compensated by a positive energy flux. 

Using the same logic, we propose another heuristic and even more ambitious interpretation of what happens after the vanishing of the exotic matter supporting an Ellis wormhole. After the negative energy is removed, the wormhole snaps and the outcome depends highly on the amount of positive energy, which shall compensate for it. As for the ER=EPR interpretation, the entanglement between the degrees of freedom in the two asymptotic regions is broken. So maybe, this is where the energy (or at least part of it) came from; the broken entanglement. Another way to phrase it is to say that the energy was hidden in the form of entanglement. Again, one could in principle compute the energy released by the entanglement between the two regions and check if it corresponds to the energy predicted by \cite{Ford:1999qv}. However, only if the energy due to the broken entanglement is bigger, this calculation can be used to disprove our heuristic picture, since it may be only part of the overall positive compensating flux. 

Finally, we should mention that throughout our discussion we carefully left an important question unanswered. Namely: \textit{what} is entangled? This is a deep question whose answer we cannot hope to justify within this simple and seemingly unrelated work. However, our educated guess is: \textit{Qubits} of a microscopic theory from which spacetime emerges. For some time, the existence of an underlined structure of spacetime has been a subject of many popular projects (\cite{VanRaamsdonk:2010pw}, \cite{Jacobson:2015hqa}). Recently, the idea that both gravity and spacetime arise from the entanglement between degrees of freedom of a microscopic theory, has been used to reproduce even observational data (\cite{Verlinde:2016toy}). In fact, in \cite{Verlinde:2016toy}, the idea that a mass corresponds to the decrease of entanglement entropy has been a central argument.   

Even though the above discussion was somewhat hand-waving, we argued that the connection between energy, entanglement and connectedness is a deep one. It is not only at the center of quantum information theory research, but it has proven to be a useful tool in our search for pieces of the puzzle that is quantum gravity.

% \subsection{Further restrictions on W/H traversability}
%\label{futher restriction}

\subsection{QFC}
\label{QFC}

The \textit{Quantum Focusing Conjecture} (QFC), as the name suggests, is a conjectured generalization of the Classical Focusing (sec. \ref{the focusing}). The statement of the conjecture is that the \textit{Quantum Expansion} $:=\Theta$ is non-increasing:

\begin{equation}
\label{QFC statement}
0\geq \Theta^{\prime} \;.
\end{equation} 

As in the case of QNEC (sec. \ref{QNEC}), the meaning of this statement can be explained in a few words, but the derivation and interpretation of the quantities involved are rather involved. In the first part of this section we shall assume the responsibility of summarizing the above.  For a more detailed treatment, we refer the reader to the original paper \cite{Bousso:2015mna}.

The setup is the same as that of sec. \ref{QNEC} and $S''_{out}$ is again the second functional derivative of the entanglement entropy of the fields sitting on the one-half of the Cauchy surface $M$. $\theta'=\frac{d\theta[\lambda(y)]}{d\lambda(y)}$ is the rate of the expansion we defined earlier in \ref{Raychaudhuri equations}, now seen also as a functional derivative.

The motivation for QFC came from studying the \textit{generalized entropy}, a quantity introduced by Bekenstein in \cite{Bekenstein:1972tm}, which is defined as the sum of all Black Hole horizons in Planck units and the entropy of the matter fields outside them.

\begin{equation}
\label{Sgen}
S_{gen} \equiv \frac{A}{4G_{N} \hbar} + S_{out}  \;.
\end{equation}

The original purpose of Bekenstein was to find an entropy that can be used to prevent violations of the \textit{Generalized Second Law} of thermodynamics. A famous example of such a violation occurs in the evaporation process of a Black hole, in which the area of the Horizon decreases due to an ingoing NEC violating flux. Computing the entanglement entropy of the radiated quanta $S''_{out}$, one can see that the entropy loss due to the area decrease is overcompensated.

One of the most remarkable features of $S_{gen}$ is that, although it involves the UV-divergent von Neumann entropy $S_{out}$, it is itself finite and cut-off independent. In \cite{Bousso:2015mna}, a mechanism through which both the leading and subleading divergences of $S_{out}$\footnote{The entanglement entropy on the one side of a sharp boundary such as $\Sigma$, is highly divergent due to short distance entanglement between field degrees of freedom very close to the boundary.} are canceled by the renormalization of $G_{N}$ in $\frac{A}{4G_{N} \hbar}$ is presented. In short, these divergences are absorbed into counterterms of the gravitational action.

To obtain the quantum expansion from the $S_{gen}$, we look a fig. \ref{QNEC figure}. Using the same logic as in QNEC, we cast the quantities involved in \eqref{Sgen} in the form of functionals:

\begin{equation}
\label{Sgen 2}
S_{gen}[\Sigma(\lambda(y))] \equiv \frac{A[\Sigma(\lambda(y))]}{4G_{N} \hbar} + S_{out}[\Sigma(\lambda(y))]  .
\end{equation}

Then $\Theta$ is defined as the first functional derivative of $S_{gen}$:

\begin{equation}
\label{Theta}
\Theta[\Sigma(y);y_1] \equiv \frac{4G\hbar}{\sqrt[]{g(y_1)}} \frac{\delta S_{gen}}{\delta \Sigma (y_1)} \equiv \lim_{{\mathcal{A}}\to0}  \frac{4G\hbar}{\mathcal{A}} \frac{\delta S_{gen}}{\delta \Sigma (y_1)}\;.
\end{equation}

Where $\sqrt[]{g(y_1)}$, is the finite area element of the metric restricted on $\sigma$.
Finally, the QFC inequality reads:

\begin{equation}
\label{QFC formula}
\frac{\delta }{\delta \Sigma (y_2)} \Theta[\Sigma(y);y_1] \leq 0\;.
\end{equation}

In words, $\Theta$ cannot increase at point $y_1$, if the slice of N defined by $\Sigma(y)$ is infinitesimally deformed along the generator $y_2$ of N in the same direction.

 Since the only important quantity here is $S_{gen}$, the regime of validity of the conjecture, coincides with the regime of validity of $S_{gen}$. Originally, $S_{gen}$ was defined for causal horizons, but following \cite{C:2013uza} \cite{Myers:2013lva} \cite{Engelhardt:2014gca} the writers assumed that $S_{gen}$ can be used to describe any co-dimension 2, Cauchy splitting surface $\Sigma$. Moreover, the surface need not be connected, nor compact.

After having formally defined QFC, it is interesting to see how one can arrive at the QNEC bound (sec. \ref{QNEC}) from it. Here it is useful to define the classical infinitesimal expansion as:

\begin{equation}
\label{QFC classical expansion}
\theta \equiv \lim_{\textbf{A}\to0}\frac{1}{\mathcal{A}}\frac{d\mathcal{A}}{d\lambda}\;,
\end{equation}

where $\mathcal{A}$ is an infinitesimal area element which separates infinitesimal neighboring null geodesics. Now combining \eqref{QFC classical expansion} and the definitions of $\Theta$ \eqref{QFC formula} and $S_{gen}$ \eqref{Sgen 2}:

\begin{equation}
\label{Theta theta}
 \Theta = \theta +\frac{4G \hbar} {\mathcal{A}}S'_{out}\;.
\end{equation}

Above and from now the limit is implicit but always there. By taking another variation at the same point, i.e. looking only at the diagonal part of QFC, one has: %++say somewhere that the off-diag part is always positive by the usage of ??strong subadditivity?? 

\begin{equation}
\label{Theta der}
\begin{split}
0\geq \Theta^{\prime} & = \theta^{\prime} +\frac{4G \hbar} {\mathcal{A}} (S_{out}^{\prime\prime} - S_{out}^{\prime} \theta)\\
& = -\frac{1}{2} \theta^2 - \sigma^2 + \omega^2 -8 \pi G_{N} \langle T_{kk} \rangle +  \frac{4G \hbar} {\mathcal{A}} (S_{out}^{\prime\prime} - S_{out}^{\prime} \theta)\;.
\end{split}
\end{equation} 

If one chooses a congruence for which $\theta=\sigma= \omega = 0$, then the above statement reduces to QNEC:

\begin{equation}
\label{QNEC from QFC}
\langle T_{kk} \rangle \geq \frac{\hbar}{2\pi \mathcal{A}} S''_{out}\;.
\end{equation}

Moreover, if one assumes that the classical limit corresponds to simply taking $\hbar \rightarrow 0$, then one obtains the classical focusing theorem from \eqref{Theta der}:

\begin{equation}
\theta' \leq 0\;,
\end{equation}

and from \eqref{QNEC from QFC}, an expression claimed to be the classical NEC of the form:

\begin{equation}
\langle T_{kk} \rangle \leq 0\;.
\end{equation}

However, we believe that going from the quantum to the classical case is more subtle than taking $\hbar \rightarrow 0$. In addition, (as we will discuss extensively in ch.\ref{Semiclassical}) $\langle T_{kk} \rangle$ is not always well defined classically. %++ask Ben if he agrees++

%Say W/H stuff at last...

\bigskip

\bigskip

%\subsubsection{No transmission Principle}
%\label{Netta}

\newpage

\section{Semi-Classical Gravity}
\label{Semiclassical}

In the current chapter, we discuss Semi-Classical Gravity and its relevance to the violations of NEC in QFT.
Semi-classical gravity is an approximation which is used when one wants to describe a theory for which matter is quantized, but gravity (spacetime) is not. There are many versions of this approximation, but most of them have the following 
equation at their core:

\begin{equation}
\label{SCG}
G_{\mu\nu}= 8\pi G_N \langle T_{\mu\nu} \rangle_\psi\;.
\end{equation}

\medskip

 In this equation, the expectation value of the quantized matter fields in some state $ \lvert \psi \rangle$ is used as a source of a classical fixed background spacetime. Now, this is not a very precise statement since, as we know from GR, there are more than one metrics that can correspond to the same stress tensor\footnote{An example is the vacuum $T_{\mu\nu}=0$, where solutions to the Einstein equations are Schwarzschild, Minkowski, etc.}. Moreover, if one tries to find one of these solutions, he would have the usual, cumbersome task of solving some very difficult partial, non-linear differential equations. In fact, this task is so difficult, that there is no good ``general'' mathematical procedure of producing exact solutions to these equations. The ones we have were mainly obtained by assuming a lot of symmetry (stationarity, spherical symmetry etc.).

However, the Einstein equation can be solved quite easily following another route. Instead of choosing a stress tensor and requesting that it solves these equations, we can start with a metric of a Lorentzian manifold and compute the corresponding stress tensor (ch. \ref{intro}). Then, the task simplifies to finding a ``physical" theory for which the expectation value of the stress tensor in one of its states, is the one predicted by the Einstein equations. Again this is not an easy task, but it is in principle easier than the previous one. This way of solving the Einstein equations solves both the problems of ``uniqueness" and difficulty, and it is the scheme people usually use to construct wormholes or other exotic solutions.

That said, another reason for which one should be worried about the semi-classical gravity equation, is that it involves a classical object $G_{\mu\nu}$, and a quantum expectation value $\langle T_{\mu\nu} \rangle_\psi$. However, in support of the validity of \eqref{SCG}, one can provide the following argument. There is a regime in which quantum effects of matter are important but quantum effects of gravity are not. For example, there clearly exists an energy scale in which quantum contributions to gravity have not yet become important, but quantum contributions to matter fields have become so important that they dominate the processes. If this were not true, it would be impossible for us to make predictions for experiments at CERN without considering quantum gravity corrections. In fact, Quantum Electrodynamics (QED), our most precise quantum theory, is an example in which quantum gravity corrections are not taken into account. Nevertheless, in the next chapter (ch. \ref{VSCG}) we shall see that things are more subtle and that semi-classical gravity is not only sensitive to the energy scale.

So we have presented a short introduction to semi-classical gravity, and now as promised we shall see its connection to NEC. In previous chapters, we noted that a lot of work has been done towards generalizing GR theorems beyond GR but also finding a way to source wormholes and other exotic spacetime solutions. However, a very important question that has not been addressed in the above literature is the following: in which regime do the above attempts take place? And the much-expected answer is the regime of semi-classical gravity. We do not yet hold enough pieces of the puzzle that is quantum gravity, to be able to even determine its shape, hence it is not strange that so much work has been done in the domain of semi-classical gravity. A reasonable claim supporting this choice is that semi-classical gravity is some limit of quantum gravity.

\subsection{Regime of validity of semi-classical gravity}
\label{VSCG}

Semi-classical gravity, like any other approximation, holds in a specific regime and as it should, breaks down if one tries to treat something that cannot be treated this regime. The regime of validity of semi-classical gravity is the set of conditions which we silently imposed when we defined it in the previous chapter. These conditions were 1) that the background metric is fixed and 2) that gravity is not quantized. 

% *maybe there are more?*

For reasons that will become clear in sec. \ref{conjecture}, we focus on when semi-classical gravity breaks down, or cannot be trusted. To our knowledge this happens in the following cases:

\begin{enumerate}

\item The semi-classical gravity approximation breaks down in high energy scales, i.e. close to Planck scale, where the full quantum gravity treatment is needed.
% +Say about Q correction becoming big cause coupling is irrelevant bla bla? Say smthing else?+

\medskip 

\item The semi-classical gravity approximation breaks down when the fluctuation induced to the metric by the fluctuations of the stress energy tensor, namely the 2-point function $\langle T_{\mu\nu} T_{\sigma\rho} \rangle_\psi $, in the state where the expectation value is taken, are large (sec. \ref{fluct}). (Even away from Planck scale)

%\medskip

%item The semi-classical gravity equation does not make sense for ``very quantum" states. (Even away from Planck scale)

\end{enumerate}

The first case is a well-known and studied break down of semi-classical gravity, and therefore we will not discuss it in great detail. However, the latter, which has received less attention, is central to our main topic. Hence, we shall devote several of the upcoming sections to its analysis (sec. \ref{fluct}, sec. \ref{Very QM}.

\subsection{Conjecture}
\label{conjecture}

Putting everything discussed in the previous two chapters (ch. \ref{Semiclassical}, sec. \ref{VSCG}) together, led us to the following question: Can semi-classical gravity be used to treat the QFT states in which \eqref{naive QNEC} is violated? We believe that the answer to this question is negative.
Following this intuition, we propose a simple, no-go conjecture:

\begin{center}
\fbox{\begin{minipage}[t]{12.5cm}
\textbf{Conjecture}: \textit{For states in which $\langle T_{\mu\nu} K^\mu K^\nu \rangle_\psi < 0$, the semiclassical \\ equation $ G_{\mu\nu}=8\pi G_N \langle T_{\mu\nu} \rangle_\psi$ is not a valid approximation.}
\end{minipage}}
\end{center}

where $\langle T_{\mu\nu} K^\mu K^\nu \rangle_\psi \geq 0 $ is the naive version of the Quantum Null Energy Condition we defined in sec. \ref{sec:The form of NEC in QFT}. If our conjecture holds, it provides us with a way out of the stalemate the violations of NEC in QFT put us in. The route is the following: if $\langle T_{\mu\nu} K^\mu K^\nu \rangle_\psi \geq 0 $ holds in semi-classical gravity, then no W/H's or other exotic spacetimes can exist and we are free to use $\langle T_{\mu\nu} K^\mu K^\nu \rangle_\psi \geq 0 $ to generalize all of the GR theorems that used NEC as an assumption to the semi-classical gravity regime. This exit exists, of course, only in the regime of semi-classical gravity, but as we discussed this is the actual regime of interest. In the following chapters, we shall see the ways in which states with $\langle T_{\mu\nu} K^\mu K^\nu \rangle_\psi < 0$, violate the conditions of semi-classical gravity.

\subsubsection{Fluctuations of the Stress Tensor}
\label{fluct}

The first to connect the NEC violating states with large fluctuations of their stress tensor were Cheng-I Kuo \& L.H. Ford \cite{Kuo:1993if}. In their paper, they propose the dimensionless quantity $\Delta$, as a measure of the fluctuations of the stress tensor, where $\Delta$ is defined us:

\begin{equation}
\label{first Delta}
\Delta_{\mu\nu\rho\sigma} = \left \lvert  \frac{ \langle :T_{\mu\nu} (x) T_{\rho\sigma} (y): \rangle_\psi - \langle :T_{\mu\nu} (x): \rangle_\psi \langle :T_{\rho\sigma}(y): \rangle_\psi}{\langle :T_{\mu\nu} (x) T_{\rho\sigma} (y): \rangle_\psi} \right \rvert\;.
\end{equation} 

\medskip

Since computing all these components requires a rather long algebraic calculation, they suggest that taking the coincident limit $x \rightarrow y$ for the $\mu=\nu=\rho=\sigma=0$ component is enough to prove their point. Thus, the quantity they actually proceed to calculate is: 

\begin{equation}
\label{delta}
\Delta (x) = \left \lvert  \frac{ \langle :T_{00} (x) T_{00} (x): \rangle_\psi - \langle :T_{00} (x): \rangle_\psi \langle :T_{00} (x): \rangle_\psi}{\langle :T_{00} (x) T_{00} (x): \rangle_\psi} \right \rvert \;.
\end{equation} 

Here $\langle :T_{\rho\sigma}(x): \rangle_\psi$ is the expectation value of the normal ordered stress tensor in some state $\lvert \psi \rangle$ and $\langle :T_{\mu\nu} (x) T_{\rho\sigma} (y): \rangle_\psi$ is the normal ordered 2-point function. As we have explained in sec. \ref{0+2}, the normal ordering procedure is a regularization which consists of putting all creation operators to the left and subtracting the VEV of the operator of interest. The fluctuations of the stress tensor are said to be small in the state of interest, if $\Delta \ll 1$ and large if $\Delta \approx 1$. We must note here that this condition is as strong as NEC since it is local in the same sense, i.e. it is imposed on each and every point of spacetime.
 
Thereafter, they proceed to calculate $\Delta$ for \textit{coherent states} and some known NEC violating states: the \textit{Casimir vacuum}, the \textit{vacuum plus two particles} and the \textit{squeezed vacuum states}. For the coherent states, they find $\Delta = 0$, which was to be expected since it is common wisdom that coherent states correspond to classical field excitations. Thus it would be highly unusual if they were incompatible with semi-classical gravity. To our knowledge, there is no well-behaved theory that contains coherent states which exhibit NEC violating behavior. For all NEC violating states mentioned above, the value of $\Delta$ is $\approx 1$ or greater.

However, there is room for doubt as to if this quantity is indeed enough to quantify the amount by which the semi-classical gravity approximation is violated. The first caveat we identify, is that the object $\langle :T_{00} (x) T_{00} (x): \rangle_\psi$ is only part of $\langle :T_{00} (x) : : T_{00} (x): \rangle_\psi$, which is the actual 2-point function at the coincident limit. Although $\langle :T_{00} (x) : : T_{00} (x): \rangle_\psi$ contains divergent pieces\footnote{In some very rare cases of states and for some specific component of the stress tensors, the two are equivalent, e.g for the photon number eigenstate: \begin{equation*} \langle :T_{00} (x) T_{00} (x): \rangle \approx \langle:T_{00} (x) : : T_{00} (x): \rangle \end{equation*} }, one should find a way of extracting the information encoded in these pieces rather than just subtracting them. In later work, Ford \& Wu realized this problem, but did not present a non-divergent, well-defined quantity that measures the fluctuations $\langle :T_{00} (x) : : T_{00} (x): \rangle_\psi$. In sec. \ref{Smear}, we present a quantity which not only solves this problem but is in principle more physical. 

\subsubsection{Very Quantum States}
\label{Very QM}

In sec. \ref{sec:The form of NEC in QFT}, we mentioned an example that makes using the expectation value of an operator in a classical equation counter-intuitive. Hereon, we will call such states \textit{very quantum states} and we will argue against their compatibility with the semi-classical approximation.

\begin{figure}[H]
\centering
\includegraphics[width=0.5\textwidth]{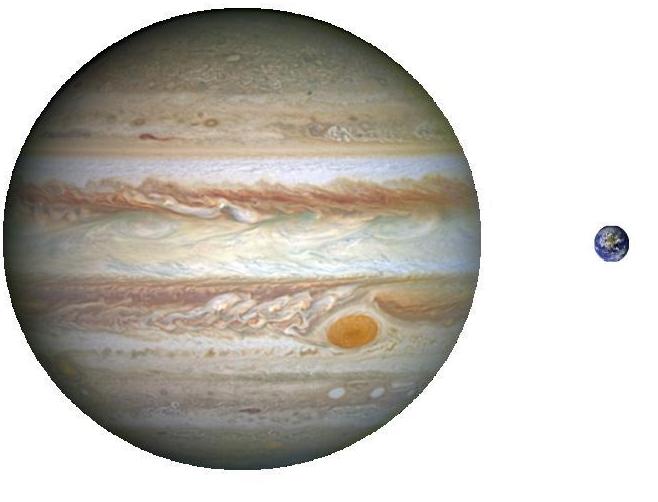}
\caption{Superposition: 50\% Jupiter ($318 M_{\oplus}$) and 50\% Earth ($1M_{\oplus}$). Expectation value is: Planet X($159.5M_{\oplus}$) }
\label{planets figure}
\end{figure}

The operator of interest is again the stress energy tensor. In the probability distribution of $T_{\mu\nu}$'s eigenvalues for very quantum states, one can find eigenvalues ``far'' from each other which carry high probability weight. Since the eigenvalues of the stress tensor are matter distributions in spacetime, if they are far apart in the probability distribution functional, they correspond to very different geometries. Quantitatively, one expects the fluctuations of $T_{\mu\nu}$ to be big and to induce equally big metric fluctuations, leading to a breakdown of the semi-classical approximation. In fig. \ref{planets figure}, we give a simplified example of such a state. The simplification refers to the fact that the eigenvalues of $T_{\mu\nu}$ describe the matter density at points, not areas. However, this last inconsistency can be restored if one uses a ``smeared'' version of  $T_{\mu\nu}$, as we shall see in sec. \ref{Smear}.

\subsubsection{The Probability distribution}
\label{prob}

At the end of the day, if one wishes to make precise statements about the expectation value of the stress tensor in some state, he needs to study the probability distribution of the eigenvalues of $\hat{T}_{\mu\nu} (x)$ in that state. The first thing that one should notice, is that each one of these eigenvalues corresponds to a different distribution of matter in spacetime. This means that by assuming the semi-classical gravity equation, each eigenvalue of $\hat{T}_{\mu\nu} (x)$ corresponds to a different spacetime\footnote{We have already discussed in ch. \ref{Semiclassical} that $T_{\mu\nu} (x)$ does not correspond to a \textit{single} solution of the Einstein or SCG equation. However this fact does not affect our argument.}, making the expectation value a sum over possible spacetimes multiplied by the probability they have to occur.

Our purpose here is to use the probability distribution function, to distinguish between semi-classical and non-semi-classical states. The most important quantity for this purpose is the variance $\sigma^2$. The variance we shall consider is $\sigma^2 = \langle T_{kk} (x) T_{kk} (x) \rangle_\psi - \langle T_{kk} (x) \rangle_\psi \langle T_{kk}(x) \rangle_\psi$, for some state $\psi$. Here, use $T_{kk} = T_{\mu\nu} K^{\mu} K^{\nu}$ since we are interested in the null behavior of $T_{\mu\nu}$ and $\sigma^2$ should be a scalar quantity. For very quantum states, the fluctuations are always bigger than the mean, leading to a big variance. On the other hand, for semi-classical states, the variance in most cases is close to zero since they should have sharply peaked around the mean probability distributions. Another useful feature of the variance is that it throws away states whose existence leads to a probability distribution of $T_{\mu\nu}$ which contains heavy tails. For such distributions, the variance is big, since the peak of the probability distribution is ``drugged way'' from the mean, due to the huge expectation values of the tail. 

Taking all the above into account  we define the following quantity as a measure of  the \textbf{magnitude of fluctuations} of $\langle T_{kk} \rangle_\psi$:

\begin{equation}
\label{Measure of Fluct's}
\frac{\sigma^2}{\langle T_{kk}(x) \rangle^2_\psi} =  \frac{ \langle T_{kk} (x) T_{kk} (x) \rangle_\psi - \langle T_{kk} (x) \rangle_\psi \langle T_{kk}(x) \rangle_\psi} {\langle T_{kk}(x) \rangle^2_\psi}\;.
\end{equation}

When $\frac{\sigma^2}{\langle T_{kk}(x) \rangle^2_\psi} << 1$ state $\psi$ is \textit{semiclassical}. When $\frac{\sigma^2}{\langle T_{kk}(x) \rangle^2_\psi} \approx1 ~\mbox{or} > 1$, the state $\psi$ is \textit{very quantum}. In addition to its other features, this is a scalar quantity and hence co-ordinate invariant. Although the quantity \eqref{Measure of Fluct's} has the required form, it contains the divergent quantity $ \langle T_{kk} (x) T_{kk} (x) \rangle_\psi$. Thus, in the next chapter we fix this problem in a way that we believe is the most natural and efficient.

\subsection{Smearing}
\label{Smear}

The most natural way to regulate the divergences in $\langle T_{kk} (x) T_{kk} (x) \rangle_\psi$ is to define a \textit{smeared} version of the stress tensor. Here \textit{smearing} means integrating the root of all divergences, operator $T_{kk} (x)$, with some smearing function $f_{x_0}(x)$ over the whole spacetime; this results in the smearing of the quantities that the stress tensor describes over some finite region controlled by the width of the smearing function $\tau$.

\begin{equation}
T^{s}_{kk} (\textbf{x}_0, \tau, t)= \int^{\infty}_{-\infty} f_{x_0}(x) T_{kk} d^dx.
\end{equation}

First let us discuss the characteristics the function should have. It should be a smooth function, and it should integrate to one: $\int^{\infty}_\infty f_{x_0}(x) dx = 1$. A function that satisfies the above and is easy to handle, is a Gaussian function of the form:

\begin{equation}
\label{smearing f}
f_{x_0}(x)=\left(\frac{1}{4\pi\tau^2}\right)^{\frac{d}{2}} e^{-\left(\frac{x_0 -x}{2\tau}\right)^2}.
\end{equation}

Where $\textbf{x}_0$ is the vector which defines the position of the center of the Gaussian in space, and  $\tau$ determines its width. Smearing does not only act as a natural \textit{cut-off} to the divergences of $\langle T_{kk} (x) \rangle_\psi$ and $\langle T_{kk} (x) T_{kk} (x) \rangle_\psi$, but also makes the planet intuition more precise. Comparing matter densities  $ T_{kk} (x) $ at points with planets does not make so much sense. However, $ T^{s}_{kk} $ describes the matter density in a \textit{region}, which makes it a much more suitable quantity for the description of stellar objects. In fact $ T^{s}_{kk} $ is in general more physical than the stress tensor at a point since this is what one expects to measure in real life experiments.

\bigskip

\subsubsection{The Smeared Stress Energy Tensor}
\label{Smeared T computation}

In this section, we define a smeared version of the stress energy tensor and calculate $\langle T_{++} (x) \rangle$ for the NEC violating state sec. \ref{0+2 state} in 1+1-D in light-cone co-ordinates. For that purpose, we make use of the smearing function \eqref{smearing f}, and similar techniques to the ones we used in sec. \ref{0+2}. One main difference here is that, instead of the circle on which we worked on in sec. \ref{0+2}, we now work on the infinite plain. Thus we are required to cast the sums of sec. \ref{0+2} into integrals, change our commutation relations accordingly and take care of some IR divergences which were previously controlled by the size of the circle. First, using the dispersion relation $w=\lvert k \rvert$ we see that the scalar modes decouple to \textit{right} and \textit{left movers}. Since we are interested in computing $\langle T_{++} (x) \rangle$ we only consider the right movers, namely:

\begin{equation}
\label{right movers}
 \phi(x_{+})= \int_{w=0}^{w=\infty} \frac{dw}{w} \left( a_w e^{-ix^+w} + a^\dagger_w e^{ix^+w}. \right)
\end{equation}

The commutation relations of the annihilation and creation operators take the following form:

\begin{equation}
\label{com a_w}
\begin{split}
[\hat{a}_k,\hat{a}^{\dagger}_{k'}] & = w_k~ \delta (k-k'), \\
[\hat{a}_k,\hat{a}_{k'}] & = 0, \\
[\hat{a}^{\dagger}_k,\hat{a}^{\dagger}_{k'}] & = 0.
\end{split} 
\end{equation}

where $\delta (k-k')$ is the Dirac delta function. The stress energy tensor is:

\begin{equation}
\label{T++}
\begin{split}
 T_{++} = \int_{w=0}^{w=\infty} dw \int_{w'=0}^{w'=\infty} dw' ( & - a_w a_{w'} e^{-ix^+(w+w')}+ a^{\dagger}_w a_{w'} e^{-ix^+(w'-w)} + \\ & + a^{\dagger}_{w'} a_w e^{-ix^+(w-w')} - a^{\dagger}_{w} a^{\dagger}_{w'} e^{ix^+(w'+w)} ).
\end{split}
\end{equation}

Here and from now on, we use the normal-ordered stress tensor, omitting the colons (:...:). The Gaussian smearing function we use, since we are interested in integrating along $x^+$-axis is:

\begin{equation}
\label{smearing f++}
f_{x^+_0}(x^+)= \left(\frac{1}{4\pi\tau^2} \right)^{\frac{1}{2}} e^{-\left(\frac{x^+_0 -x^+}{2\tau}\right)^2}.
\end{equation}

The integral over $x^+$ with integrand $T_{++}$ and $f_{x^+_0}(x^+)$ is simple Gaussian integral which yields:

\begin{equation}
\label{Smearing one}
\begin{split}
T^{s}_{++} (\textbf{x}^+_0, \tau, x^+) & = 
\int_{w=0}^{w=\infty} dw \int_{w'=0}^{w'=\infty} dw' ( - a_w a_{w'} e^{-ix^+_0(w+w')-\tau^2(w+w')^2} + \\ &+ a^{\dagger}_w a_{w'} e^{-ix^+_0(w'-w)-\tau^2(w-w')^2} + a^{\dagger}_{w'} a_w e^{-ix^+_0(w-w')-\tau^2(w-w')^2}- \\ & - a^{\dagger}_{w} a^{\dagger}_{w'} e^{ix^+(w'+w)-\tau^2(w+w')^2} ) .
\end{split}
\end{equation}

Hereafter, we take \eqref{Smearing one} as our definition for the smeared stress tensor operator.

\subsubsection{Fluctuations of Smeared Stress Tensor}
\label{fluctuations}

The next object we need to compute in order to obtain the variance of the stress tensor, are the fluctuations of the stress tensor. Much like the original, not normal ordered $T_{++}$ the UV divergences of $T_{++}T_{++}$ are due to its VEV. Thus, the vacuum state is the perfect candidate for testing the validity of our claim that the $\langle T^s_{++}T^s_{++} \rangle_\psi$ converges. Taking $\langle 0 \rvert  T^s_{++}T^s_{++} \lvert 0 \rangle$ one finds that only one term contributes, namely:

\begin{equation}
\label{TT0 surv term}
\begin{split}
\langle 0 \rvert  T^s_{++}T^s_{++} \lvert 0 \rangle & = \int_{w_i=0}^{w_i=\infty} \prod_{i=1}^{4} dw_i  \cdot \langle 0 \rvert a_{w_1} a_{w_2} a^{\dagger}_{w_3} a^{\dagger}_{w_4} \lvert 0 \rangle \cdot \\ & \cdot  e^{-ix^+_0(w_1+w_2-w_3-w_4)-\tau^2[(w_1+w_2)^2+(w_3+w_4)^2]} \\ & =
\int_{w_1=0}^{w_1=\infty} \int_{w_2=0}^{w_2=\infty} dw_1 dw_2 \cdot  w_1 w_2 \cdot \\ & \cdot  2e^{-2\tau^2(w_1+w_2)^2} \\&
= \frac{1}{\tau^4} \cdot 4.2 \cdot 10^{-2}   .
\end{split}
\end{equation}

We see that \eqref{TT0 surv term} is indeed finite, and depends only on our cut-off. Now that we have obtained a finite $\langle T^s_{++}T^s_{++} \rangle_\psi$ we upgrade our \textit{measure of semi-classicality} of states: 

\begin{equation}
\label{Measure of Fluct's rev}
\boxed{\frac{\sigma^2}{\langle T^s_{kk}(x) \rangle^2_\psi} =  \frac{ \langle T^s_{kk} (x) T^s_{kk} (x) \rangle_\psi - \langle T^s_{kk}(x) \rangle^2_\psi} {\langle T^s_{kk}(x) \rangle^2_\psi}}
\end{equation}
\medskip

and finally use it to test our conjecture.

\subsubsection{Testing the Conjecture}
\label{test conj}

In this section, we present our results for the fluctuations of three representative classes of states. We first consider the fluctuation of the very quantum \textit{vacuum}, which we find to be very big in accordance with our discussion. Next, we compute our measure of semi-classicality for the NEC violating 0+2 particles state. We find that the quantity of interest \eqref{Measure of Fluct's rev} is again $>>1$; a result which supports our conjecture (sec. \ref{conjecture}). Finally, we calculate \eqref{Measure of Fluct's rev} for a class of \textit{coherent states}, which we find to be $<<1$. Coherent states are semi-classical which, to the best of our knowledge, do not violate NEC. Hence, this last result tests our definition of semi-classicality and provides additional evidence that supports the validity of \ref{conjecture}.

As promised, we first compute \eqref{Measure of Fluct's rev} and measure the \textit{quantumness} of the vacuum. Since the variance of the vacuum state is:
 
 \begin{equation}
\label{Vacuum variance}
\begin{split}
\sigma^2_0 = \langle 0 \rvert  T^s_{++}T^s_{++} \lvert 0 \rangle - \langle 0 \rvert  T^s_{++} \lvert 0 \rangle^2 = \frac{1}{\tau^4} \cdot 4.2 \cdot 10^{-2}.   
\end{split}
\end{equation}

We trivially obtain:

\begin{equation}
\label{Vacuum Fluct's}
\frac{\sigma^2_0}{\langle T^s_{kk}(x) \rangle^2_0} = \infty >> 1.
\end{equation}

Thus our result agrees with the generally accepted idea that the vacuum is a very quantum state. However, the vacuum does not violate NEC and thus the above result, although interesting, does not validate or falsify our conjecture. Hence, we turn to the simple 0+2-particles state we considered in sec. \ref{0+2}. The only subtlety here is that if we naively consider the exact same state as in sec. \ref{0+2} some IR divergences will appear. That is because in the ``continuous limit'' we consider for our calculation in the current chapter (\ref{Smear}), the state $\lvert 2_k \rangle$ is not properly normalized. Taking into account the commutation relations \eqref{com a_w}, we choose the following normalization:

\begin{equation}
\label{normalization}
\lvert 2_k \rangle = \frac{a^{\dagger}_k a^{\dagger}_k}{\sqrt[]{2}~w_k~ L} \lvert 0_k \rangle .
\end{equation}

Using the above, we compute the expectation value of our stress energy tensor.

\begin{equation}
\label{0+2 smeared stress}
\begin{split}
\langle  T^s_{++} \rangle_{0+2}=\frac{w_k}{L} \left(1-\sqrt[]{\frac{3}{2}} \cos(2x^+_0 \cdot w_k) e^{-4\tau^2w^2_k} \right).
\end{split}
\end{equation}
 
This result is similar to sec. \ref{0+2 SET}, but here due to the smearing, the negative contribution on the r.h.s of eq.\eqref{0+2 smeared stress} is exponentially damped. Since the width $\tau$ controls the area of spacetime in which our source is contained, eq.\eqref{0+2 smeared stress} has an interpretation familiar to us from QIs and ANEC. The bigger the spacetime region we are considering, the smaller the NEC violating flux.    

To obtain the expression for the fluctuations of the stress tensor in the 0+2-particles state, one needs to go through lengthy wick contractions and perform some simple Gaussian integrals. After following these algebraic steps, one has:

\begin{equation}
\label{0+2 smeared fluctuations}
\begin{split}
\langle  T^s_{++}T^s_{++} \rangle_{0+2} &=  \frac{1}{L} \Big(\frac{w_k}{\tau^2}e^{-2\tau^2w^2_k} -  \frac{w^2_k}{\tau} ~  \sqrt[]{\frac{\pi}{2}} \cdot \mbox{Erfc}(\sqrt[]{2}\tau w_k) +  \\ &+ \frac{w^2_k}{\tau} \cdot \sqrt[]{\frac{\pi}{2}} \cdot (1+\mbox{Erf}(\sqrt[]{2}\tau w_k) \big) - \\ & - 1.225\frac{w^2_k}{ \tau}\cdot e^{-2\tau^2w^2_k} ~ \cos(2x^+_0\cdot w_k) \Big) + \\ & + \frac{w^2_k}{L^2} \Big(2 + e^{-4\tau^2w^2_k} \Big)
+\frac{0.0625}{\tau^4} \;,
\end{split}
\end{equation}

and thus the variance is:

\begin{equation}
\label{0+2 smeared variance}
\begin{split}
\sigma^2_{0+2} &=   \frac{1}{L} \Big(\frac{w_k}{\tau^2}e^{-2\tau^2w^2_k} -  \frac{w^2_k}{\tau} ~  \sqrt[]{\frac{\pi}{2}} \cdot \mbox{Erfc}(\sqrt[]{2}\tau w_k) +  \\ &+ \frac{w^2_k}{\tau} \cdot \sqrt[]{\frac{\pi}{2}} \cdot (1+\mbox{Erf}(\sqrt[]{2}\tau w_k) \big) - \\ & - 1.225\frac{w^2_k}{ \tau}\cdot e^{-2\tau^2w^2_k} ~ \cos(2x^+_0\cdot w_k) \Big) + \\ & + \frac{w^2_k}{L^2} \Big(2 + e^{-4\tau^2w^2_k} \Big)
+\frac{0.0625}{\tau^4}  - \\ & - \frac{w^2_k}{L^2} (1-\sqrt[]{\frac{3}{2}} \cos(2x^+_0 \cdot w_k) e^{-4\tau^2w^2_k} )^2.
\end{split}
\end{equation}

Before computing the measure of semi-classicality for this state there are two factors we should take into account. The first and most relevant one is that $w_k L >>1$. Since we have taken the continuum limit, we secretly assumed that the density of states is high enough to do so. Thus, $w_k$ must be sufficiently big and since L the size of our space the dimensionless $w_k L >>1$ must indeed be large. So, it is sensible to write the measure of fluctuations in a way that clarifies which terms are relevant, and which are not:

\begin{equation}
\label{0+2 flucts}
\begin{split}
\frac{\sigma^2_{0+2}}{\langle  T^s_{++} \rangle^2_{0+2}} &=  \Big(\frac{w_k L}{\tau^2}e^{-2\tau^2w^2_k} -  \frac{w^2_k L}{\tau} ~  \sqrt[]{\frac{\pi}{2}} \cdot \mbox{Erfc}(\sqrt[]{2}\tau w_k) +  \\ &+ \frac{w^2_k L}{\tau} \cdot \sqrt[]{\frac{\pi}{2}} \cdot (1+\mbox{Erf}(\sqrt[]{2}\tau w_k) \big) - \\ & - 1.225\frac{w^2_k L}{ \tau}\cdot e^{-2\tau^2w^2_k} ~ \cos(2x^+_0\cdot w_k) + \\ & + w^2_k \Big(2 + e^{-4\tau^2w^2_k} \Big)
+\frac{0.0625 L^2}{\tau^4}  - \\ & - w^2_k (1-\sqrt[]{\frac{3}{2}} \cos(2x^+_0 \cdot w_k) e^{-4\tau^2w^2_k} )^2 \Big) / \\ &
~~ w^2_k (1-\sqrt[]{\frac{3}{2}} \cos(2x^+_0 \cdot w_k) e^{-2\tau^2w^2_k} )^2.
\end{split}
\end{equation}

If one looks carefully (and/or treats this expression numerically), he will notice that for all values of $\tau$ the fluctuations are indeed $>>1$. Nevertheless, we take the interesting limit $w_k \rightarrow \frac{1}{\sigma}$ which gives a clear, analytic result:

\begin{equation}
\label{0+2 flucts simplified}
\begin{split}
\frac{\sigma^2_{0+2}}{\langle  T^s_{++} \rangle^2_{0+2}} &=  \Big(w_k L~e^{-2} - w_k L ~  \sqrt[]{\frac{\pi}{2}} \cdot \mbox{Erfc}(\sqrt[]{2}) +  \\ &+ w_k L \cdot \sqrt[]{\frac{\pi}{2}} \cdot (1+\mbox{Erf}(\sqrt[]{2}) \big) - \\ & - 1.225~ w_k L\cdot e^{-2} ~ \cos(2x^+_0\cdot w_k) + \\ & + \Big(2 + e^{-4} \Big)
+0.0625~ w^2_k L^2  - \\ & - (1-\sqrt[]{\frac{3}{2}} \cos(2x^+_0 \cdot w_k) e^{-4} )^2 \Big) / \\ & ~~ (1-\sqrt[]{\frac{3}{2}} \cos(2x^+_0 \cdot w_k) e^{-2} )^2 \approx \\ & \approx \frac{0.0625~ w^2_k L^2}{(1-\sqrt[]{\frac{3}{2}} \cos(2x^+_0 \cdot w_k) e^{-2} )^2}  + \mathcal{O}(w_k~L) >>1. \\
\end{split}
\end{equation}

\medskip

It crucial that we also present an example for which the measure of fluctuations (\eqref{Measure of Fluct's rev}) becomes small ($>>1$). According to our conjecture, this should be a semi-classical state, and as we have discussed the most natural candidates are coherent states. To make our life easier we consider a state which is a superposition of many particles of a single mode. Following Roy J. Glauber \cite{Glauber}  we use the \textit{displacement operator} $D_k(z)$ acting on the vacuum to create the above mention states.

\begin{equation}
\label{coherent}
\lvert \psi \rangle_{coh} \equiv D_k(z) \lvert 0 \rangle .
\end{equation}

Where 

\begin{equation}
\label{dis operator}
D_k(z) \equiv e^{ (z~a^{\dagger}_k -z^{*}~a_k) } = e^{-\frac{\lvert z \rvert^2}{2}w_k \delta(0)} ~ e^{z~a^{\dagger}_k} ~ e^{-z^{*}~a_k} ,
\end{equation}

and $z= s~e^{i\gamma}$. In \eqref{dis operator} use made use of the \textit{Baker–Campbell–Hausdorff formula} in order compute the commutation relations we need. Moreover, from this final form of $D_k(z)$ in \eqref{dis operator}, one can immediately observe that the state we obtained is indeed a 1-mode excited coherent state since the action of $e^{-z^{*}~a_k}$ on the vacuum reduces to that of the identity. 

The commutation relations we will need are the following:

\begin{equation}
\label{comm coh}
\begin{split}
[a_{k},a^{\dagger}_{k'}] & =  \delta (k-k')~w_k\;,  \\
[a_{k},D_{k'}] & =  z~ D_{k'}~ \delta (k-k')~w_k \;, \\
[D^{\dagger}_{k'},a^{\dagger}_{k}] & = z^{*}~ D^{\dagger}_{k'}~ \delta (k-k')~w_k  \;. \\ 
\end{split}
\end{equation}

After performing similar calculations to the above two cases, we obtain the following expressions for the expectation value of the stress tensor and its fluctuations:

\begin{equation}
\label{stress coh}
\begin{split}
\langle T^{s}_{++} \rangle_{coh}=2~w^2_k~s^2 \left[1-\cos2(\gamma -x^+_0~w_k) ~ e^{-4\tau^2w^2_k} \right]\;,
\end{split}
\end{equation}

\begin{equation}
\label{fluct coh}
\begin{split}
\langle T^{s}_{++} T^{s}_{++} \rangle_{coh} & = \frac{0.042}{\tau^4} + 2~w^4_k~ s^4~e^{-8\tau^2w^2_k} + 2~w^4_k~ s^4~e^{-8\tau^2w^2_k} \cos(4\gamma-x^+_0 w_k) \\ & + 4S^4 w^4_k - 8S^4 w^4_k\cos2(\gamma-x^+_0 w_k) e^{-4\tau^2w^2_k} - \\ & -
\frac{2~w^2_k~ s^2~e^{-2\tau^2w^2_k}}{\tau^2}\cos2(\gamma-x^+_0 w_k)+ 2s^2 \frac{w^2_k}{\tau^2} e^{-2\tau^2w^2_k}+ \\ & + \frac{2 w^3_k~ s^2 ~e^{-2\tau^2w^2_k}}{\tau}~\sqrt[]{\frac{\pi}{2}}~\left(1+\mbox{Erf} \left(\sqrt[]{2}~w_k~\tau \right)-\mbox{Erfc}\left(\sqrt[]{2}~w_k~\tau \right)\right)\;. \\
\end{split}
\end{equation}

It is worth noticing that the quantity $\langle T^{s}_{++} \rangle_{coh}$ is always positive \eqref{stress coh}, and thus the semi-classical coherent states do not violate the quantum version of NEC. All that is left, is to compute the measure of the stress tensor fluctuations, and show that it is indeed $<<1$.

\begin{equation}
\label{coh flucts}
\begin{split}
\frac{\sigma^2_{coh}}{\langle  T^s_{++} \rangle^2_{coh}} &= \frac{ \langle T^{s}_{++} T^{s}_{++} \rangle_{coh} -\langle T^{s}_{++} \rangle^2_{coh}}{\langle T^{s}_{++} \rangle^2_{coh}} = \\ & = \Big( \frac{0.042}{\tau^4} + 2~w^4_k~ s^4~e^{-8\tau^2w^2_k} + 2~w^4_k~ s^4~e^{-8\tau^2w^2_k} \cos 4 (\gamma-x^+_0 w_k) \\ & + 4s^4 w^4_k - 8s^4 w^4_k\cos2(\gamma-x^+_0 w_k) e^{-4\tau^2w^2_k} - \\ & -
\frac{2~w^2_k~ s^2~e^{-2\tau^2w^2_k}}{\tau^2}\cos2(\gamma-x^+_0 w_k)+ 2s^2 \frac{w^2_k}{\tau^2} e^{-2\tau^2w^2_k}+ \\ & + \frac{2 w^3_k~ s^2 ~e^{-2\tau^2w^2_k}}{\tau}~\sqrt[]{\frac{\pi}{2}}~\left(1+\mbox{Erf} \left(\sqrt[]{2}~w_k~\tau \right)-\mbox{Erfc}\left(\sqrt[]{2}~w_k~\tau \right)\right)- \\ & - 4~w^4_k~s^4 \left[1-\cos2(\gamma -x^+_0~w_k) ~ e^{-4\tau^2w^2_k} \right]^2 \Big)  / \langle T^{s}_{++} \rangle^2_{coh}\;. \\ 
\end{split}
\end{equation}

To analyze the above expression, we need to understand the role of $s$. It is easy to understand from \eqref{dis operator} that $s$ is connected to the number of single mode excitations of our state, and hence to obtain a coherent state $s$ should be big. %order N??
For $s=0$, one gets back the vacuum and the more $s$ grows, the more coherent the state becomes. Even without taking this into account, the exponential damping terms may seem enough make these fluctuations small. Nevertheless, one should also check the interesting limit where these damping terms reduce to $\mathcal{O}(1)$ factors; namely the limit $w_k \rightarrow \frac{1}{\tau}$. Remembering that $s$ is big, and noticing that it appears in the denominator of \eqref{coh flucts} as $\mathcal{O}(s^4)$ we keep only terms of $\mathcal{O}(s^4)$ in the numerator, after having taken the limit of interest. After taking into account the above considerations, we obtain the following form for the fluctuations measure in a coherent state:

\begin{equation}
\label{coh flucts limit}
\begin{split}
\frac{\sigma^2_{coh}}{\langle  T^s_{++} \rangle^2_{coh}} & = \Big( 2~w^4_k~ s^4~e^{-8} + 2~w^4_k~ s^4~e^{-8} \cos 4 (\gamma-x^+_0 w_k) \\ & + 4s^4 w^4_k - 8s^4 w^4_k\cos2(\gamma-x^+_0 w_k) e^{-4} - \\ &  - 4~w^4_k~s^4 \left[1-\cos2(\gamma -x^+_0~w_k) ~ e^{-4} \right]^2 \Big) / \langle T^{s}_{++} \rangle^2_{coh} = \\ & = 0 (<<1) \;.\\ 
\end{split}
\end{equation}

Hence, even in this extreme limit, the fluctuations are negligible. The above result provides us with evidence to support our claim that for semi-classical states the fluctuations of the stress tensor are small, and subsequently also support our conjecture (sec. \ref{conjecture}).

\subsubsection{Proposed Bound}
\label{proposal}

As we have discussed extensively in this chapter (ch. \ref{Semiclassical}), the existence of states that violate a quantum null energy condition is connected to the features of the probability distribution of the stress energy tensor's eigenvalues in these states. To first order, this reduces to the study of the said distribution's variance $\sigma^2$. In this spirit, we propose a quantum-corrected QFT bound which takes into account the relation between the violations of NEC in some state and $\sigma^2$. It is a semi-local bound, which we believe should hold for \textit{all}, not just semi-classical, states. 

\medskip

\textbf{Bound}:
\begin{equation}
\label{bound}
\boxed{ \langle T^{s}_{kk} \rangle_\psi + \sigma_\psi \geq 0}
\end{equation}

\medskip
 
 Here $\sigma$ is the standard deviation and $\langle T^{s}_{kk} \rangle$ is the usual expectation value of the null-null component of the smeared stress tensor, in some state $\psi$. It is important to note that our bound holds not only for the Gaussian function but for any smearing function satisfying the characteristics described in sec. \ref{Smear}. We have not proven the above bound, thus it still remains a proposal. Nevertheless, we have yet to see a class of states which violates it.
 
 Just as in the case of our conjecture, to provide evidence that our bound holds, we present our results for the \textit{vacuum}, 0+2 particles state (sec. \ref{0+2}) and the \textit{coherent state}
\medskip

For the vacuum, since $\langle T^s_{kk} \rangle_0=0$, the bound reduced to the following simple formula:

\begin{equation}
\label{Bound for vacuum}
\sigma_0 =~ \sqrt[]{\frac{0.042}{\tau^4}} \geq 0.
\end{equation}

Next, we present the example of the $\lvert \psi \rangle = \frac{\sqrt[]{3}}{2} \rvert 0_k \rangle + \frac{1}{2} \rvert 2_k \rangle $ state. Again here we consider the 1+1-D case, where $T^{s}_{kk} \equiv T^{s}_{++}$. Remembering our computation in \eqref{0+2 flucts simplified}, we notice that the contribution of the $\langle T^{s}_{++}\rangle^2_{0+2}$ term is of sub-leading order in $w_k~L$, in fact it is of $\mathcal{O}(1)$. Thus, $\sigma$ is of $\mathcal{O}(w_k~L)$ and $\langle T^{s}_{++}\rangle_{0+2}$ is again of $\mathcal{O}(1)$ and our bound is once again satisfied trivially.

\begin{equation}
\label{Bound for 0+2}
\begin{split}
\sigma_{0+2} + \langle T^{s}_{++}\rangle_{0+2} \approx \sigma_{0+2}  \geq 0.
\end{split}
\end{equation}

Finally, again in the relevant limit $w_k \rightarrow \frac{1}{\sigma}$, since the variance for our class of coherent states is zero, the only contribution to the bound comes from the, always positive, stress tensor:

\begin{equation}
\label{Bound coh}
\begin{split}
\sigma_{coh} + \langle T^{s}_{++}\rangle_{coh} =  2~w^2_k~s^2 \left[1-\cos2(\gamma -x^+_0~w_k) ~ e^{-4\tau^2w^2_k} \right]  \geq 0.
\end{split}
\end{equation}

\newpage

\section{Conclusions}
\label{conclusions}

The simple quantum version of NEC is violated in QFT (sec. \ref{naive QNEC}), but as we saw in sec. \ref{perturbed} it is harder to violate than expected. In many cases, the NEC violating states (sec. \ref{sec:1+1 CFT}, sec. \ref{0+2}, squeezed vacuum states etc.) are superpositions, whose interference takes the form of an oscillatory term responsible for the violation. However, not all representatives of these classes violate NEC. Whether the oscillatory term leads to violations of NEC or not, depends on the normalization of the state; an example where this becomes apparent is the 0+2 particles states, for which exactly half of the phase space covered by these states gives rise to NEC violations and the other half does not \cite{Pfenning:1998ua}.
An interesting feature of these states is that they satisfy ANEC, making them a good candidate for exotic matter sources, in contrast to states (sec. \ref{Casimir}, moving mirrors, Hawking radiation etc.) which do not contain such oscillatory terms and usually lead to more severe violations.

 A local version of NEC in QFT is most likely not very useful. It is interesting that a state dependent, NEC-like, local bound can be proven for certain QFT's. However, the bound itself is not strict enough, and cannot be ``tuned'' to provide us with information about the non-local scale at which the NEC violating contribution become negligible. As we have discussed, the more local the bound, the greater the amount of exotic matter it allows for. We strongly believe that for a NEC-like bound to make sense in QFT, it should be semi-local. Hence, we have proposed a bound for a smeared version of the stress energy tensor, which is (much like QI's) semi-local. In addition, this result is unique since it is the only bound that takes into account characteristic features of the probability distribution of the stress tensor. Moreover, it has a very simple interpretation in terms of our conjecture: since the variance of the stress tensor for NEC violating states is large, the negative contribution due to $\langle T^s_{kk} \rangle_\psi$ should be compensated by the standard deviation. However, this raises the question of whether or not these contributions get overcompensated, making our bound too strong. Hence before worrying about proving this proposed bound, we must check which one of the existing bounds is the sharpest one.

 So far, there exists no bound in QFT that allows us to generalize all theorems of GR and exclude Wormholes and other exotic spacetimes. QNEC (sec. \ref{QNEC}) is a local QFT bound that does not restrict the amount by which NEC is violated enough to do the above. The semi-local QI's are doing a much better job at that, but there exist solutions in literature which manage to avoid them \cite{Visser:2003yf}. A similar statement can be made about ANEC, which is indeed the most restrictive bound we have, but its power is severely weakened by its strictly global nature. The only way we know of to salvage everything, at least at the semi-classical level, is for our conjecture (sec. \ref{conjecture}) to be true. 
 
 However, we should not fail to mention that one can think of a natural counter-example to this conjecture, which is the Hawking radiation. Two essential points of the black hole evaporation process are that it takes place in the semi-classical regime and that NEC is violated. Nevertheless, in \cite{Wu:1999ea} Wu and Ford, using their quantity $\Delta$, calculated that the fluctuations of the Hawking flux to be big. Whether or not the Hawking radiation contradicts our conjecture is an interesting question that we would like to think about in the future. 
 
It is true that some constructs that seem to shape our classical reality are not inherited by the quantum one (e.g. determinism). In the end, semi-classical gravity has one foot in the classical world and one in the quantum, and it may very well be that not everything we like in GR will also work in semi-classical gravity. 

\bigskip

\bigskip

\textbf{Acknowledgments} I am grateful for conversations with Vassilis Anagiannis, Mikola Schlottke, Isaías Roldán, Vittorio Ricciardulli, Yaroslav Shustrov, and Erik van Heusden. I would like to thank Diego Hofman and Jan Pieter van der Schaar for their advise and for stimulating discussions. Finally, I would like to thank my supervisor Ben Freivogel whose intuition and deep thought guided me throughout this project.

%\subsection{Future goals}
%\label{goals}

%Prove the no-go conjecture:

%\textbf{Conjecture}: \textit{For states in which $\langle T_{\mu\nu} K^\mu K^\nu \rangle_\psi < 0$  the semiclassical equation $ G_{\mu\nu}=8\pi G_N \langle T_{\mu\nu} \rangle_\psi$ is not a valid approximation} 
%bigskip

%One of the few things that exclude exotic spacetimes \& allows the generalization of theorem \& excludes Wormholes and the other exotic spacetimes in SCG
%\medskip

%Find the sharpest bound that holds for $\langle T_{\mu\nu} K^\mu K^\nu \rangle$.

%\medskip

%Study extensively the probability distribution functional. Hard, but the best way to see if one should trust expectation values of $T_{\mu\nu}$

\end{document}